\documentclass[letterpaper,twocolumn,10pt]{article}
\usepackage{usenix}

\usepackage{xspace}
\usepackage[ruled]{algorithm2e}
\SetAlCapNameFnt{\small}
\SetAlCapFnt{\small}
\usepackage{soul}
\usepackage{adjustbox}
\usepackage[]{hyperref}
\usepackage{comment}
\usepackage{cprotect}
\usepackage{multirow}
\usepackage{tabularx}
\usepackage{xcolor}
\usepackage{xurl}

\usepackage{enumitem}
\usepackage{listings}
\usepackage{subcaption}
\usepackage{makecell}

\usepackage{censor}
\usepackage{float}
\usepackage{amsmath}
\usepackage{makecell}
\usepackage{soulutf8}
\usepackage{fancyvrb,newverbs}
\usepackage{booktabs}
\usepackage{anyfontsize}
\usepackage{hhline}
\usepackage{boldline}
\usepackage{array}

\usepackage{arydshln}
\usepackage{float}
\usepackage{cprotect}
\makeatletter
\g@addto@macro{\endtabular}{\rowfont{}}
\makeatother
\newcommand{\rowfonttype}{}
\newcommand{\rowfont}[1]{
\gdef\rowfonttype{#1}#1\ignorespaces%
}
\makeatother
\colorlet{soulhl}{yellow!40}
\sethlcolor{soulhl}

\newcommand{\diffbg}[3][0pt]{{\fboxsep#1\colorbox{#2}{\strut #3}}}
\newcommand{\niparagraph}[1]{\vspace{2pt}\noindent\emph{\textbf{#1}}}
\def\backtick{\`{}}
\lstdefinelanguage{code}{
  basicstyle=\fontsize{8}{8}\selectfont\ttfamily,
  backgroundcolor=\color{lightgray!8},
  morecomment=[f][\diffbg{red!20}]-,
  morecomment=[f][\diffbg{green!20}]+,
  morecomment=[f][\textit]{@@},
  literate={`}{\backtick}1 {■}{\rule{5pt}{5pt}}1,
}

\lstdefinelanguage{inlinecode}{
  basicstyle=\ttfamily,
}
\lstdefinelanguage{prompt}{
  basicstyle=\fontsize{8}{8}\selectfont\sffamily,
  backgroundcolor=\color{cyan!8},
  breaklines=true,
  breakindent=0pt,
  tabsize=1,
  showlines=true,
  gobble=8pt, 
  extendedchars=true,
  columns=flexible,
  language={},
  literate={`}{\backtick}1 {■}{\rule{5pt}{5pt}}1, 
}
\lstdefinelanguage{llm}{
  basicstyle=\fontsize{8}{8}\selectfont\sffamily,
  backgroundcolor=\color{red!8},
  breaklines=true,
  breakindent=0pt,
  tabsize=1,
  showlines=true,
  gobble=8pt, 
  extendedchars=true,
  columns=flexible,
  language={},
  literate={`}{\backtick}1 {■}{\rule{5pt}{5pt}}1, 
}

\newcommand{\name}{ECO\xspace}
\newcommand{\namexplain}{\underline{\textbf{E}}ngine for \underline{\textbf{C}}ode \underline{\textbf{O}}ptimization}

\newcommand{\code}[1]{\lstinline[language=inlinecode]{#1}\xspace}

\newcommand{\google}{{Google\xspace}}
\newcommand{\linesofcode}{billions of lines of code}

\newcommand{\lstPrompt}{
\renewcommand{\lstlistingname}{Prompt}
}
\newcommand{\lstResponse}{
\renewcommand{\lstlistingname}{Response}
}
\newcommand{\lstCode}{
\renewcommand{\lstlistingname}{Code Snippet}
}
\newcommand{\lstCodeDiff}{
\renewcommand{\lstlistingname}{Code Diff}
}

\newcommand{\totalLinesOfCodeChanged}{25k\xspace}
\newcommand{\totalCommits}{6.4k\xspace}
\newcommand{\numberOfFuncs}{10M\xspace}
\newcommand{\revertRate}{<0.5\%}
\newcommand{\numberOfAntipatternsInTraining}{55k\xspace}
\author{
  {\rm Hannah Lin\thanks{Hannah Lin and Martin Maas contributed equally to this work.}$^{*1}$\quad
  Martin Maas$^{*1}$\quad
  Maximilian Roquemore\thanks{Now at Point72, work done entirely at Google.}$^{\dagger 2}$\quad
  Arman Hasanzadeh$^1$\quad
  Fred Lewis$^2$} \\ 
  {\rm Yusuf Simonson$^2$\quad
  Ameya Shringi$^2$\quad
  Hongwen Dai$^2$\quad
  Patrick Musau$^1$\quad
  Tzu-Wei Yang$^2$
  } \\
  {\rm 
  Amir Yazdanbakhsh$^1$\quad
  Deniz Altinbüken$^1$ \quad
  Florin Papa$^2$\quad
  Maggie Nolan Edmonds$^2$\quad
  Aditya Patil$^2$} \\
  {\rm
  Don Schwarz$^2$\quad
  Satish Chandra\thanks{Now at Meta, work done entirely at Google.}$^{\ddagger 2}$\quad
  Chris Kennelly$^2$\quad
  Milad Hashemi$^1$\quad
  Parthasarathy Ranganathan$^2$}\\
  $^1$Google DeepMind, $^2$Google
}

\begin{document}
\title{ECO: An AI-Driven Code Efficiency Optimizer for Warehouse Scale Computers (Operational Systems)}

\maketitle

\begin{abstract}
Large Language Models (LLMs) have shown significant promise in automating code efficiency optimization. While prior work demonstrates these techniques on artificial datasets such as programming competitions or small benchmarks, deploying these techniques at scale in production has remained an open problem. Arguably, two challenges have prevented the adoption in large-scale real-world systems: opportunity localization and reliability. First, applying an LLM to every line across a large code base is expensive and prone to generating an overwhelming number of low-quality suggestions, placing unsustainable cognitive load on human code reviewers. Second, the inherent unreliability of LLM-generated code risks introducing errors that can lead to production incidents. These challenges are largely orthogonal to the ML techniques prior work has focused on; they are real-world systems problems.

This paper introduces \name, a system that automatically modifies source code to improve performance at scale. \name overcomes the localization problem by combining fleet-wide continuous profiling to identify performance-critical code with an embedding-based search to pinpoint specific optimization candidates, guided by a mined dictionary of performance anti-patterns. It overcomes the reliability problem through a multi-stage verification approach that uses automated testing, LLM-based self-review, and post-deployment monitoring to ensure changes are both correct and effective. Fully productionized and deployed within \google’s hyperscale production fleet, \name has successfully landed over 6,400 commits, changing more than 25,000 lines of production code. Incorrect changes are caught before they are submitted to production, and 99.5\% of the submitted commits did not cause any rollbacks. These optimizations have resulted in savings equivalent to several hundred thousand normalized CPU cores, showing that \name makes LLM-based optimization both practical at scale and highly impactful in real-world settings.
\end{abstract}

\section{Introduction}
\label{sec:introduction}

Over the past few years, an increasing amount of work has shown the potential of Large Language Models (LLMs) to analyze and optimize code for performance \cite{gong2025languagemodelscodeoptimization,shypula2024learning,berger2023scalene,puri2021codenet,chen2022learning,chen2023supersonic,10.5555/3737916.3738283,liu2024evaluating,du2024mercury,pmlr-v267-huang25as}. By automating a task that traditionally requires extensive and costly manual engineering, LLMs promise to unlock significant efficiency gains, particularly in the long tail of workloads. A study of Google's data centers has shown that a large fraction of CPU cycles in warehouse-scale computers are spent on workloads that are individually too small to justify manual optimization effort \cite{10.1145/2749469.2750392}. Automating this process could therefore yield substantial resource and energy savings.

However, while these LLM-based approaches show promise on benchmarks such as programming competitions \cite{shypula2024learning,chen2022learning,agakov2006iterativeoptimization,garg2022deepdev,10.5555/3737916.3738283,du2024mercury}, directly applying them in a large-scale, production environment presents significant challenges that hinder their practical adoption. The challenges are not with the LLMs alone (where existing work has focused), but with making them practical in the context of real-world systems:

\begin{itemize}
  \item \textbf{Opportunity Localization}: LLM-based methods need to be provided the code to optimize. In production environments with millions or billions of lines of code, naively running an LLM across the entire repository is computationally prohibitive and generates an overwhelming number of false positives. A practical system must first find and localize high-impact optimization opportunities before attempting to apply a fix.
  \item \textbf{Reliability}: State-of-the-art LLM methods often report success as the best result out of many attempts (e.g., top-$\mathcal{K}$ accuracy) \cite{garg2022deepdev}. In a real-world production setting, however, precision is paramount. Any incorrect change can cause outages, and the high rate of subtly buggy or non-performant suggestions from an unguided LLM creates a large verification burden.
\end{itemize}

\noindent In addition, benchmarks fail to capture \textbf{system integration} and \textbf{attribution}. In complex, large-scale systems, attributing a performance inefficiency to a specific piece of code is difficult \cite{10.1145/3694715.3695977}. Furthermore, AI-based optimization must integrate with existing engineering workflows, including testing, code review, and deployment, aspects absent in benchmarks.

We present \name (\namexplain), a system that makes LLM-based optimization practical at scale by introducing novel techniques for targeted localization and robust verification. \name{} begins by mining historical commits to create a database of performance anti-patterns. These anti-patterns are then used to guide a large-scale vector similarity search \cite{guo2020scann} across the codebase. To focus this search, we combine it with data from a fleet-wide continuous profiling system, allowing \name to identify code that is both costly and a likely candidate for a known optimization.

Once a candidate is localized, \name{} uses a fine-tuned LLM to generate a code transformation. To ensure the reliability of this edit, it is passed through a multi-stage verification pipeline that includes automated build and unit testing, LLM-based self-review to detect common errors, and finally, human review. Once deployed, the change is continuously monitored to measure its performance impact and avoid regressions.

This paper makes the following contributions:

\begin{figure*}
    \centering
    \includegraphics[width=0.9\linewidth]{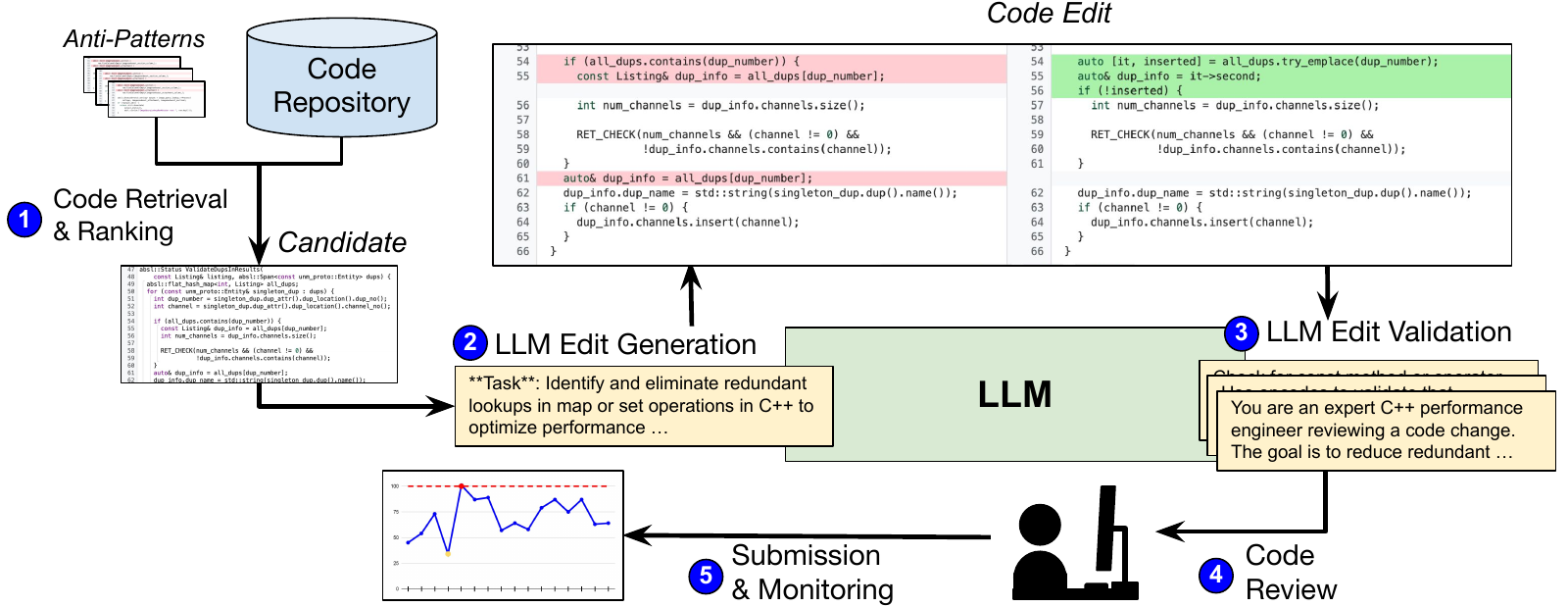}
    \caption{An end-to-end example of \name{}, demonstrating the cycle from anti-pattern mining to deployed fix.}
    \label{fig:overview:endtoend}
\end{figure*}

\begin{enumerate}
  \item An approach for identifying and localizing performance optimization opportunities by combining continuous profiling with an embedding-based similarity search guided by a mined database of anti-patterns.
  \item A methodology for making unreliable LLM-generated edits reliable for production use through automated testing, self-review, and post-deployment monitoring.
  \item The design, implementation, and evaluation of \name, a fully deployed system that embodies these techniques and successfully integrates AI-driven changes into a large-scale, human-in-the-loop engineering workflow.
  \item A study focusing on human adoption of code edits, including a human evaluation of 960 code edits to assess the quality of changes, and an at-scale production study of how human engineers adopt \name{}-generated changes in Google's production setting.
  \item An analysis of \name{}'s deployment at Google, showing it has successfully landed over 6,400 commits, changing over 25k lines of code and saving the equivalent of several hundred thousand normalized CPU cores. By filtering out and discarding incorrect changes before they are submitted to production, >99.5\% of the 6,400 commits did not have to be reverted.
\end{enumerate}

\noindent While there has been a substantial amount of prior work on LLM-based code optimization, \name{} is largely orthogonal to this work. Instead, it solves the systems problems required to turn these capabilities into scalable impact in a large-scale system infrastructure deployment. We believe it is the first operational system to achieve this.

\section{Background \& Overview}

The use of LLMs in code optimization has seen a large amount of research in the past several years \cite{gong2025languagemodelscodeoptimization}. We begin with an overview of how our approach improves upon the state of the art to solve problems that have prevented adoption of automated optimization in real-world systems.

\subsection{Why LLMs for Code Optimization?}
\label{sec:background:why_llms}

Automated code optimization is well-studied, but existing approaches have limitations. Compiler optimizations must be provably correct, preventing them from making semantic changes that a programmer knows are safe in a given context. Static analysis tools like Clang-Tidy \cite{clangtidy}, Facebook Infer \cite{harmim2019scalable}, or Semantic Patch \cite{254428} can apply rule-based edits to perform such changes, but struggle with the high syntactic variability of real-world code. These tools require narrowly defined patterns and often fail on complex control flow. For example, Clang-Tidy can suggest adding \begin{small}\verb|vector<T>::reserve|\end{small} to a simple, single-statement loop \cite{clangtidy-vector}, but cannot handle common real-world cases where a vector is populated across multiple loops, conditional branches, or helper functions (Figure~\ref{fig:dataset:anti-patterns:c}). Writing heuristics for every variant is often more work than fixing the instances manually. Further, many optimizations require domain knowledge that static tools lack, such as when a sorted map can be replaced with a faster, unsorted one.

LLMs overcome these limitations in two ways. First, they can generalize from a few examples to recognize and fix anti-patterns across a wide range of syntactic variations, reducing the need for hand-crafted heuristics. Second, they can often infer developer intent and domain-specific context from the surrounding code and comments, enabling semantic optimizations that are out of reach for traditional tools. Figure~\ref{fig:dataset:anti-patterns} shows several examples. While LLMs are more computationally expensive than static checkers, we show in Section~\ref{sec:lifecycle} that the cost of inference is negligible compared to the fleet-wide savings they unlock.

\subsection{High-Level Approach}

Figure~\ref{fig:overview:endtoend} shows an overview of \name's approach. We start by mining decades of code changes across Google's billions of lines of code spanning over 100,000 packages and thousands of projects \cite{christopher2025instructionsetmigrationwarehouse}, to identify a set of canonical code optimizations. Each consists of an \emph{anti-pattern} and a corresponding \emph{optimization} that fixes it. Two examples\footnote{Throughout the paper, we rename sensitive variables, names, or values.} of these anti-patterns, pre-sizing a vector at allocation time and removing redundant accesses to data structures, are shown in Figure \ref{fig:dataset:anti-patterns}. We group similar anti-patterns together into categories. Table~\ref{table:edit_categories} shows several of the anti-pattern categories in our database.

Using this anti-pattern database, we identify candidate locations in the codebase for similar optimizations. This problem is challenging, as exact matching fails to surface these locations due to differences in variable names and code structure. We find this problem is well-suited to machine learning and develop a code similarity search technique based on vector similarity~\cite{guo2020scann} to identify a set of \emph{candidates} similar to a given anti-pattern. To focus this search, we leverage fleet-wide continuous profiling to rank candidates based on resource consumption, identifying the best opportunities.

Subsequently, we use a fine-tuned LLM to generate code that implements the optimization. To mitigate the risk of LLM hallucinations, we improve \name's reliability by using diverse prompting strategies and a multi-stage verification pipeline. To ensure correctness, we use the LLM to self-review its generated code, similar to \cite{chen2023teaching}, run automated tests, and perform checks tailored to each project. We then submit the generated changes for human code review. After submission, we monitor the change in production to detect regressions and measure its performance impact.

\begin{table}[t]
\captionof{table}{
Anti-Pattern Categories}
\label{table:edit_categories}
\footnotesize
\begin{tabular}{p{0.15\linewidth}| p{0.6\linewidth} | p{0.1\linewidth}}
\toprule
\textbf{Category} & \textbf{Description} & \textbf{Count}\\
\midrule
Alloc & Unnecessary allocations: Allocation of a new object when reuse of an object is possible. & 8 \\
\hline
Args & Unnecessary copied args: Using an API causing unnecessary copies of arguments from a widely used internal framework. & 2501 \\
\hline
Copy & Unnecessary copies: Replacing variables with references. & 2873 \\
\hline
Map & Redundant map operations: See Figure~\ref{fig:dataset:anti-patterns:b}. & 985 \\
\hline
Move & Missing \verb|std::move|s: The last use of an object is copied instead of moved. & 1487 \\
\hline
Sort & Unnecessary sorts: Use of a sorted or stable map/set when not required. & 91525 \\
\hline
Vector & Missing vector reserves: See Figure~\ref{fig:dataset:anti-patterns:c}. & 331 \\
\bottomrule
\end{tabular}
\end{table}

\begin{figure*}[ht]
    \centering
    \includegraphics[width=1\linewidth]{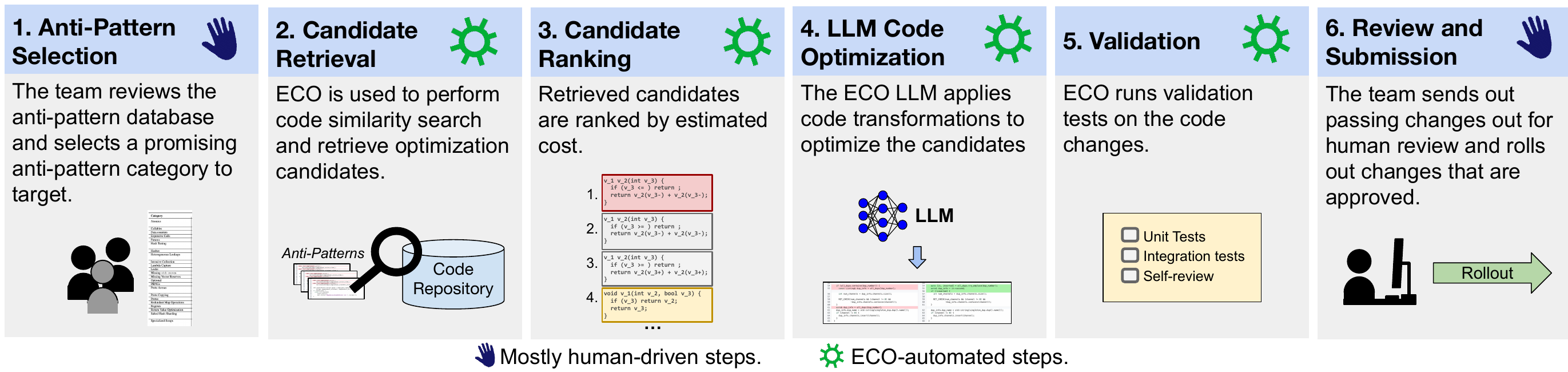}
    \caption{Operational timeline of \name{}.}
    \label{fig:overview:timeline}
\end{figure*}

\begin{figure}[ht]
    \centering
        \begin{subfigure}{\linewidth}
\begin{lstlisting}[language=code]
- bool new_entry = false;
- if (s.current_shard.find(name) == s.current_shard.end()) {
-   s.current_shard[name] = 1;
-   new_entry = true;
- } else {
-   ++s.current_shard[name];
- }
+ int64_t value = ++s.current_shard[name];
+ bool new_entry = (value == 1);
\end{lstlisting}
        \caption{\textbf{Redundant Map Operations}: Performing duplicated key lookups and reinitializing map values can waste CPU cycles. Here, the LLM had to detect that {\tt []} may add a missing key, and that for a map with integer values, it gets initialized to 0; the optimization only works in that case and is not sound in general.}
        \label{fig:dataset:anti-patterns:b}
        \end{subfigure}
        \begin{subfigure}{\linewidth}
\begin{lstlisting}[language=code]
+ result.reserve(reader->NumRecords());
  for (;;) {
    Info info;
    bool has_msg = reader->ReadToMessage(&info).ValueOrDie();
    if (has_msg) {
      result.push_back(info);
    } else {
      break;
    }
  }
  return result;
\end{lstlisting}
        \vspace{2pt}
\begin{lstlisting}[language=code]
+ error_list.reserve(3);
  int total_error_count = 0;
  for (int i = 0; i < vec1.size(); ++i) {
    const double diff = fabs(vec1[i] - vec2[i]);
    if (diff > abs_error) {
      total_error_count++;
      if (error_list.size() < 3) {
        error_list.push_back(Error(i, diff));
      }
    }
  }
\end{lstlisting}
        \cprotect\caption{\textbf{Missing Vector Reserves}:
        Vectors are a common data structure but can cause unnecessary memory reallocations that pre-allocating space can avoid. In the first case, {\tt reader} is a file API and the LLM had to determine that it can query the full number of entries with {\tt NumRecords}. It also needed the domain-specific context that {\tt has\_msg} is rarely false. In the second example, it had to conclude that the list will only have up to 3 entries.}
        \label{fig:dataset:anti-patterns:c}
        \end{subfigure}
    \caption{Examples of \name{} performance optimizations.}
    \label{fig:dataset:anti-patterns}
\end{figure}

\name{}'s optimization discovery approach results in a database of anti-patterns that can grow over time. Figure~\ref{fig:overview:timeline} outlines the end-to-end process from an operational point-of-view. A small team of engineers operates \name{}, beginning by selecting anti-patterns to target. The team ranks known anti-patterns according to observed frequency and savings potential, supported by the data and statistics gathered in the creation of the database. After selecting anti-patterns, they launch \name{} to automatically identify optimization candidates from existing examples. The team ranks these candidates (described in Section~\ref{sec:localization}), and uses a library of prompt recipes (evaluated in Section~\ref{sec:eval}) to have \name{} automatically apply an optimization across all identified locations, submitting the resulting commits for code review.

Actions by \name{} are automated, with engineers directing the system and intervening only when necessary. Most human involvement appears during the review and submission phase. However, this is required for any change generated at Google, regardless of whether or not it is AI-generated. Code changes fail if performance regressions appear in production and they need to be rolled back. \name{}'s validation process has kept the rate of rollbacks extremely low (Section~\ref{sec:lifecycle}). As such, \name{} greatly scales up human productivity compared to manually searching for optimization opportunities, writing patches and shepherding them through code review.

\section{Performance Optimization Datasets}
\label{sec:anti-patterns}

We now describe how we mine historical commits to create a dataset of efficiency anti-patterns and optimizations.

\subsection{A Dataset of Efficiency Anti-Patterns}

To enable \name{} to drive interpretable and consistent changes, we begin by identifying and categorizing performance anti-patterns from historical data. \google’s codebase contains \linesofcode{}. It is structured as a monorepo (similar to~\cite{potvin2016google, harry2017microsoft, ordogh2018twitter, lin2020uber}), but the approach would be the same if it was spread across smaller, disparate GitHub repos. While this work targets C++, \name{} is not fundamentally limited to it. Despite being organized as a monorepo, this setup covers over 100,000 packages and a wide variety of projects \cite{christopher2025instructionsetmigrationwarehouse}, including mirrors of open-source repositories \cite{thirdparty}. This diversity and scale demonstrates \name{}'s scalability and generalizability across many different kinds of projects.

All code in the repository is searchable, with a complete history of all past commits, including messages, code review comments, and code changes. This is the same setup that a typical organization or open source project building on top of GitHub would encounter. It lets us implement textual search techniques to find historical commits that improved performance. We use two search types: First, we search commit messages for keywords indicating performance improvements (e.g., "speed up", benchmark results), performance-related hashtags, and links to performance-related documentation. Second, we search curated sources like internal newsletters that publicize efficiency improvements. This process is guided by human experts who curate the anti-pattern categories, define relevant search keywords, and set up the data processing pipelines. While our coding culture encourages single-issue commits, \name{} is robust to commits with multiple changes.

We run this data extraction on a large-scale distributed computation framework, akin to Apache Beam \cite{apachebeam2012}, and maintain a database of these performance-related optimizations and the anti-patterns that they solved. We then use it as the foundation of our optimization approach. Appendix~\ref{sec:supplement:anti-patterns} provides a full list of anti-patterns in our dataset.

\subsection{LLM Fine-Tuning Datasets}
\label{sec:fine-tuning}

In addition to using our anti-pattern dataset to identify optimization candidates (Section~\ref{sec:localization}), we also combine it with other data sources to fine-tune an LLM to improve its performance on these specific code refactorings~\cite{kini2026customizingllmenterprisesoftware}. For the examples and evaluation provided in this paper, we use \name{} with Gemini Pro 1.0~\cite{team2023gemini}, which supports fine-tuning via Google's Vertex AI service~\cite{geminituning}. Our fine-tuning dataset is highly proprietary and includes source code, selected code commits, metadata of code and commits, and the performance-specific dataset derived from our anti-pattern database ($\approx$ \numberOfAntipatternsInTraining commits). The fine-tuning mechanisms are similar to prior work~\cite{petros2023didact,roziere2024code}.

While we describe a particular LLM here, \name{}'s architecture is agnostic to the specific model. We have regularly updated our model and continued to use \name{} with more recently released LLMs. However, for consistency, all evaluations in this paper were performed with the original Gemini Pro 1.0 model described above. We note that it is not the goal of this paper to evaluate the capabilities of LLMs at producing code efficiency optimizations, and models continue to rapidly improve in this area, which \name{} directly benefits from.
\section{Optimization Opportunity Localization}
\label{sec:localization}

While our anti-patterns dataset contains examples of \emph{past} optimizations, a key challenge is identifying \emph{new} candidates across our codebase. Naively applying an LLM to all of \google{}'s code is computationally expensive with an observed high false-positive rate. In addition, incremental, localized changes are favored to minimize global disruption and facilitate testing and review processes. Therefore, we introduce a search technique based on vector similarity search~\cite{guo2020scann} to identify new potential instances of anti-patterns.

\subsection{Performance IR}
\label{sec:tree}

To search for optimization opportunities, we first design an intermediate representation (IR) of functions across our codebase. We operate at the function level, as functions provide a stable and practical unit of code to track and compare across the millions of daily changes in our repository. This granularity strikes a balance between providing sufficient context for optimization and maintaining tractability. Using the Apache Beam framework~\cite{apachebeam2012}, we crawl C++ code, parse each file with Clang, and divide it into functions. We annotate each function with its AST-derived types (e.g., of arguments and local variables), which are good indicators for certain optimizations. Figure~\ref{fig:ir-example} shows an example. To avoid processing millions of functions that are not performance-critical, we filter this set to focus only on \emph{costly} functions.

\begin{figure}[t]
    \centering
\begin{lstlisting}[language=code]
{"definition": 
  "string GetTimeLabel(time_t timestamp) {
      int year, month;
      getYearMonth(timestamp, &year, &month);
      string label;
      bool success = toLabel(year, &label);
      if (!success) return getError(year, month);
      return label;
  }",
...
"variables": [
  { "name": "year", "type": "int" },
  { "name": "month", "type": "int" },
  { "name": "label", "type": "string" },
  { "name": "success", "type": "bool" }
],
"performance": [
  { "instructions": 43052 },
  { "cycles": 109572 },
  { "llc_misses": 501 }
]}
\end{lstlisting}
\caption{A performance-annotated function in the dataset.}
\label{fig:ir-example}
\end{figure}

\subsubsection{Continuous Profiling}
\label{sec:background:prof}

To identify costly functions, we leverage \emph{Google Wide Profiling} (GWP), a fleet-wide continuous profiling framework that periodically samples performance metrics from all running applications~\cite{ren2010gwp}.
For each sample, it collects the program counter, stack trace, and performance counters. These profiles are aggregated into a global database. We use this data to identify costly functions as optimization targets. In addition, correlating it with code versions allows us to evaluate changes in resource usage linked to specific commits. This helps us both identify past optimizations and also measure the impact of \name{}'s commits after deployment.

\subsubsection{Code Annotations with Performance Metrics}
\label{sec:localization:perfmetrics}

To find good optimization candidates, we annotate our IR with performance metrics such as CPU usage, memory allocations, and LLC misses from our profiler. A naive attribution of CPU usage is often unhelpful, as cycles are frequently spent in common, \emph{shared library functions} (e.g., \begin{small}\code{vector<T>::push_back}\end{small}) rather than \emph{application-specific code}. These shared libraries are not generally profitable targets for optimization at \name{}'s level, as they are usually already heavily optimized. Therefore, we re-attribute resource usage to the parent functions in the call stack that are most meaningful for optimization.

\subsubsection{Identifying Costly Functions}
\label{sec:localization:identify}

\begin{algorithm}[t]
\scriptsize
\caption{Retrieving costly functions.}
\label{fig:pruning-pseudocode}
\KwIn{A function $F$ and its parent function $P$}
\KwOut{Costly functions under $F$'s call tree}
\SetKwData{Callee}{callee}
\SetKwData{Costlies}{costly\_fns}
\SetKwData{Up}{up}\SetKwFunction{FPrune}{ShouldPrune}
\SetKwFunction{FGetCostlyFns}{GetCostlyFns}
\SetKwFunction{FGetCallees}{GetCallees}
\SetKwProg{Fn}{Function}{}{}
\BlankLine
\Fn{\FGetCostlyFns{$F$, $P$}}{
  \uIf{\FGetCallees{$F$} is empty} {
      \lIf{\FPrune{$F$, $P$}} {
        \KwRet{[]}
      }
    \KwRet{[$F$]} \;
  }
  \BlankLine
  \Costlies $\leftarrow$ []\;
  \For{\Callee in \FGetCallees{$F$}}{
    extend \Costlies with \FGetCostlyFns{\Callee, $F$} \;
  }
  \BlankLine
  \lIf{\Costlies is empty and not \FPrune{$F$, $P$}} {
    \KwRet{[$F$]}
  }
 \KwRet \Costlies \;
}

\SetKwFunction{FIsShared}{isSharedFn}
\SetKwFunction{FGetCycles}{getCyclesOfTree}
\SetKwProg{Fn}{Function}{}{}
\BlankLine
\Fn{\FPrune{$F$, $P$}}{
  \lIf{\FGetCycles{$P$} $>$ $C_{max}$}{ \KwRet {false}}
  \lIf{\FIsShared{$F$}} {
    \KwRet{true}
  } \lIf{\FGetCycles{$F$} $<$ $C_{min}$ or \FGetCycles{$F$} $>$ $C_{max}$} {
    \KwRet{true}
  }
  \KwRet {false} 
}
\end{algorithm}
\begin{figure}[t]
    \centering
    \includegraphics[width=0.8\linewidth]{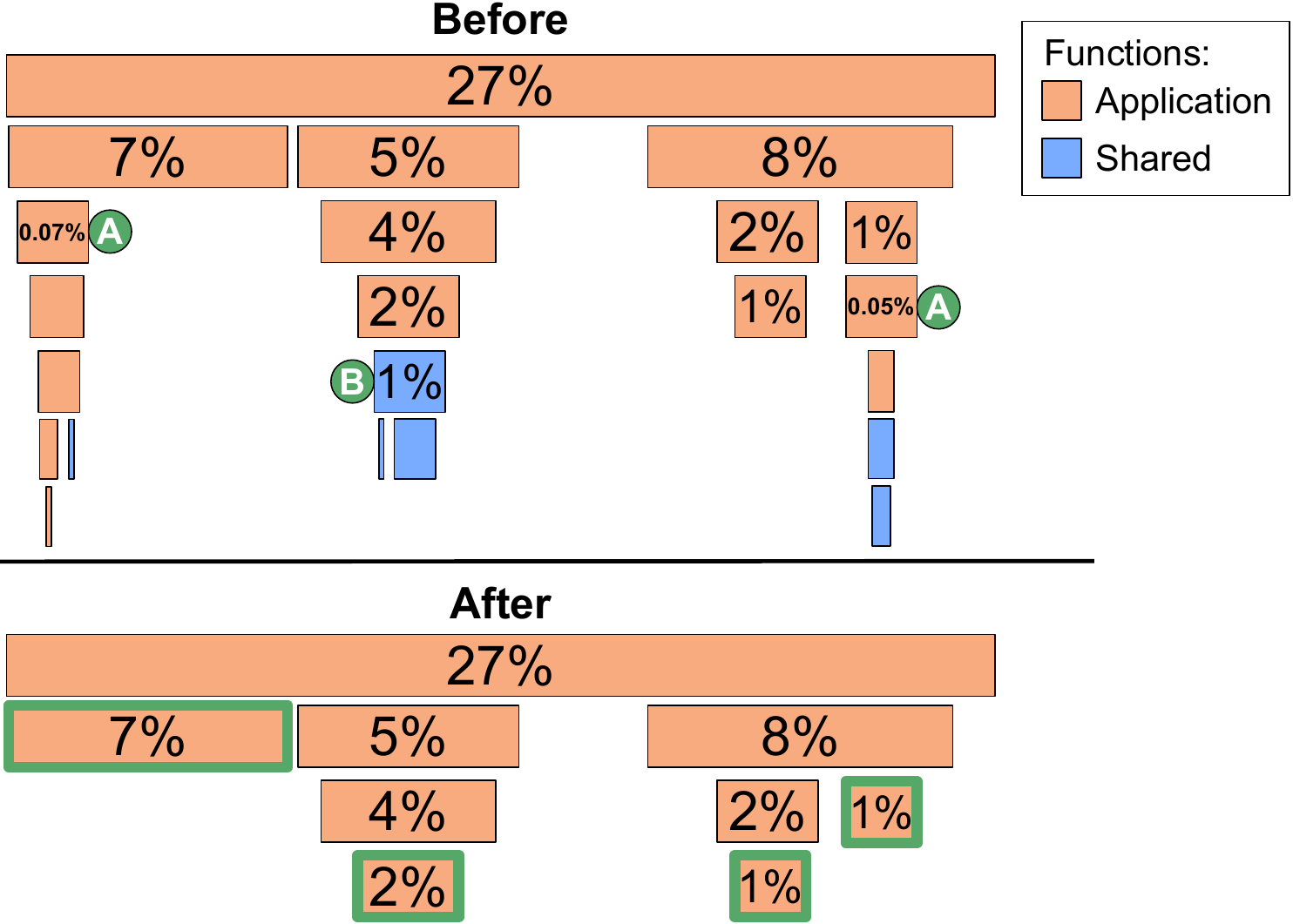}
    \caption{Pruning function call graphs (annotated by percent of the application's total cycles and represented as a flame graph \cite{gregg2016flame}, not drawn to scale). Unlabeled functions consume less than $0.01\%$ of the application's cycles. Orange represents application-specific code, blue represents shared code. \textbf{A} labels pruned sub-trees roots that form less than $C_{min}$=0.1\% of the app. \textbf{B} labels the root of a pruned shared code sub-tree. Boxes outlined green after pruning (bottom) are the leaf nodes of the pruned call graph targeted for optimization.}
    \label{fig:stack-truncation}
\end{figure}

\begin{figure*}[t]
    \centering
    \includegraphics[width=\linewidth]{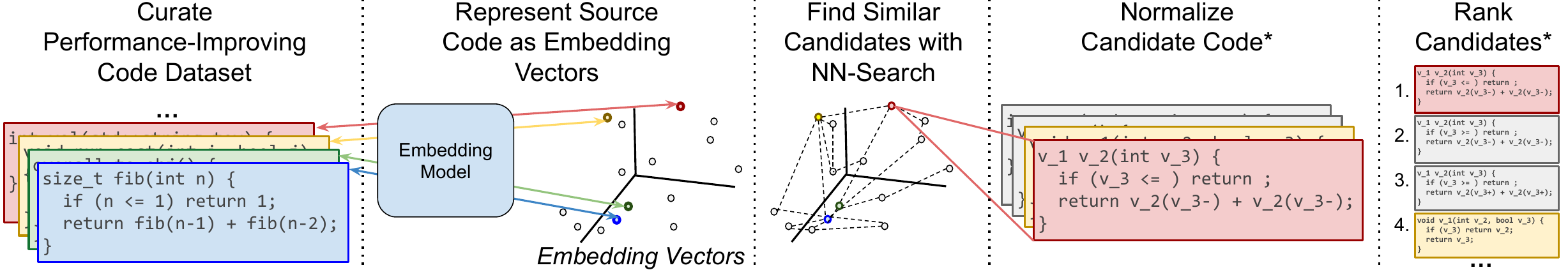}
    \caption{\name's code retrieval approach. Source code is represented as vectors and ScaNN~\cite{guo2020scann} is used to find nearest neighbors in the vector space. After retrieving nearest neighbors, we optionally rank code by similarity; these steps are marked by (*).}
    \label{fig:localization:code-sim}
\end{figure*}

We take per-binary function call trees from profiler samples and apply a pruning and attribution step, outlined in Algorithm~\ref{fig:pruning-pseudocode} and depicted in Figure~\ref{fig:stack-truncation}. We traverse the call trees to find the lowest-node \emph{application-specific functions} that are responsible for significant cost. We define application-specific functions as those common to fewer than a threshold of binaries. We recursively prune sub-trees and attribute their cost to their caller. Note that this has no effect on the caller's code representation and only affects its accounted resource consumption for the purpose of identifying costly functions. A sub-tree of shared functions is pruned and its cost attributed upwards, unless the caller is already too large (exceeding $C_{max}$ of total binary cycles). Application-specific sub-trees are also pruned upwards until a caller accounts for at least $C_{min}$ and at most $C_{max}$ of the binary's cycles. The upper threshold prevents aggregation at too high a level (e.g., a binary's main function), which is unproductive for localized optimizations. Through empirical tuning, we set $C_{min}$ to 0.1\% and $C_{max}$ to 25\%. This analysis yields over \numberOfFuncs{} potential optimization candidates, which form our \emph{search space}.

\subsection{Code Retrieval}
\label{sec:localization:code-sim}

Given our anti-pattern dictionary and a search space of costly functions, we need to find all candidates that likely match a particular anti-pattern. This is challenging, as instances of an anti-pattern can be textually and structurally diverse (e.g., Figure~\ref{fig:dataset:anti-patterns:b}). Manual heuristics like regular expressions or AST-based techniques such as CodeQL \cite{youn2023declarative} cannot usually capture this diversity. We therefore use a code retrieval technique that discovers new instances of an anti-pattern from past examples.

We first form a database of vector representations (embeddings) of each function in our search space (Figure~\ref{fig:localization:code-sim}). We then take examples from our anti-patterns database and look for similar functions in our search space using approximate nearest neighbor (ANN) search, a method proven effective in other code recommendation contexts~\cite{luan2019aroma}. We use ScaNN~\cite{guo2020scann}, an open-source vector similarity search library developed at Google, for this task. Given a vector representation of a \emph{query} (the "before" function state or a code diff embodying an anti-pattern), ScaNN returns an approximate set of similar vectors from the database. To focus on high-impact changes, we only search within the set of costly functions identified previously. For each query, we retrieve the top-$\mathcal{K}$ ($\mathcal{K}=500$) candidates. We then normalize the query and candidates by removing custom names, static strings, and comments, and rank the candidates by a \emph{syntactic similarity} score (Section~\ref{sec:localization:code-sim-ranking}).

\subsubsection{Embedding Design for Semantic Similarity}
\label{sec:localization:code-sim-embedding}

We explore two strategies for vector embeddings:

\niparagraph{\textit{(a) Bag of Words (BOW).}} This approach involves creating embeddings based on a BOW representation. Our intermediate representations provide textual tokens for each function after parsing. The embedding vector of a function is created by forming a token-frequency vector, excluding comments, common keywords (e.g.,  \verb|return|), and punctuation tokens.

\niparagraph{\textit{(b) Deep Embedding Models.}} We use a 1B-parameter dual-encoder \emph{deep text embedding} model \cite{lee2024gecko}, originally trained on English query-document pairs, and fine-tune it on 25M samples of: (1) semantically similar function pairs and (2) code diffs and their corresponding "before" function states. Fine-tuning on similar functions allows us to query for similar code snippets. Fine-tuning on diffs allows us to search for specific optimization patterns, focusing the query on the relevant lines of code rather than the entire function. This matches how we later use the deep embedding model (querying it using both code snippets and code diffs) and is similar to how others use diffs or Git commits for training models~\cite{li2023starcodersourceyou, diffcoder, li2022automating}.

\subsubsection{Ranking Performance Opportunities}\label{sec:localization:code-sim-ranking}

\setlength{\belowdisplayskip}{2pt} \setlength{\belowdisplayshortskip}{2pt}
\setlength{\abovedisplayskip}{2pt} \setlength{\abovedisplayshortskip}{2pt}

After retrieving the approximate top candidates through ANN, we use a \emph{syntactic similarity score} for further, more detailed ranking. The score $\mathcal{S}$ between a query $\mathcal{Q}$ and a candidate $\mathcal{C}$ is defined as (higher is better):
\begin{equation*}
\mathcal{S}(\mathcal{Q}, \mathcal{C}) = \frac{1}{4} [B(\mathcal{Q}, \mathcal{C}) + R(\mathcal{Q}, \mathcal{C}) + T(\mathcal{Q}, \mathcal{C}) + F(\mathcal{Q}, \mathcal{C})]
\end{equation*}
\noindent Each component of $\mathcal{S}$ is from 0 to 1 and defined as follows:

\begin{itemize}[leftmargin=*]
    \item $B(\mathcal{Q}, \mathcal{C}), R(\mathcal{Q}, \mathcal{C})$: BLEU~\cite{papineni2002bleu} and ROUGE-L~\cite{lin2004rouge} score are text similarity metrics. $B$ and $R$ are calculated after normalizing the code ($\mathcal{Q}$ and $\mathcal{C}$) to remove custom names, static strings, and comments.
    \item $T(\mathcal{Q}, \mathcal{C}) = (|T_Q \cap T_C|) / max(|T_Q \cup T_C|, 1)$, where $T_Q$, $T_C$ are types present in $\mathcal{Q}$ and $\mathcal{C}$ from our code representation (Section~\ref{sec:localization:perfmetrics}).
    \item $F(\mathcal{Q}, \mathcal{C})$: Cosine similarity of control flow-related keywords (\emph{e.g.}, \begin{small}\code{for}\end{small}, \begin{small}\code{while}\end{small}) in the functions' BOW vectors.
\end{itemize}

\noindent{}The inclusion of $B$ and $R$ captures textual similarity, while $T$ and $F$ incorporate information about types and control flow, improving the representation of semantic similarity and helping further refine the set of candidates, after the approximate match provided by ANN.
\section{Code Edit Generation Strategies}
\label{sec:edit-gen}

After identifying candidates, we generate code transformations by prompting an LLM (Section~\ref{sec:fine-tuning}). Since generation quality is highly dependent on the prompt, we explore a variety of prompting strategies, or \emph{prompting recipes}, for engineers using \name{} to build upon~\cite{wei2022chain, yao2022react}.

LLM prompting is a well-studied area~\cite{hou2024promptsurvey}. The simplest approach, \emph{zero-shot prompting}, directly instructs the model to perform a task without giving it any examples. This works for simple edits but struggles with complexity. Since token generation is when the model applies logic, zero-shot prompting leaves few steps for the model to apply complex operations. It also relies on the LLM correctly interpreting the instructions.

Several techniques address this. \emph{Few-shot} prompting includes examples of the desired input-output behavior (e.g., before/after code pairs) in the prompt to help the model with pattern matching, a technique shown to be effective for code~\cite{nashid2023fewshot}. \emph{Chain-of-Thought} (CoT) prompting instructs the model to first output a plan or reasoning steps before generating the final output~\cite{wei2022chain}. This encourages more structured reasoning and works well for code transformation~\cite{li2024cot}. Finally, \emph{agentic} ReAct-style prompting enables multi-step reasoning by having the model cycle through generating a thought, an action (e.g., a command to execute in an external environment such as, in our case, a shell), and then feeding the output of the command back into the model, allowing it to interactively refine its work~\cite{yao2022react}. The prompting recipes below apply these paradigms to \name{}'s code editing task.

\subsection{Prompting Recipes}

\begin{figure*}[ht]
\begin{subfigure}[b]{0.2\linewidth}
    \centering
\begin{lstlisting}[language=prompt]
cat { target path }
```
{ target snippet }
```
Optimize the usage of maps.

We can accumulate results and use one-time initialization to avoid unnecessary extra work.
...

Optimize the code.

patch { target path }
@@




\end{lstlisting}
      \caption{Zero-shot}
      \label{fig:prompt:db:ex}
\end{subfigure}%
\hfill
\begin{subfigure}[b]{0.15\linewidth}
\begin{lstlisting}[language=prompt]
Remove redundant map lookups to improve performance.

Example 1
{ example snippet 1 }
@@
{ example diff 1 }

Example 2
{ example snippet 2 }
@@
{ example diff 2 }

Example 3
{ target snippet }
@@


\end{lstlisting}
  \caption{Few-shot}
  \label{fig:prompt:fs}
\end{subfigure}%
\hfill
\begin{subfigure}[b]{0.3\linewidth}
    \centering
\begin{lstlisting}[language=prompt]
# Original Code
{ target snippet }

# Instructions
1. Walk through how to optimize this code using this format:
* Step 1: { step 1 }
* Step 2: { step 2 }
* Step 3: { step 3 }
2. Then provide a code diff optimizing the code using this format:
patch { target path }
@@
{ code diff }

# Answer
1. Let's think this through, step by step:
* Step 1:

\end{lstlisting}
    \caption{Chain-of-Thought}
    \label{fig:prompt:cot}
\end{subfigure}%
\hfill
\begin{subfigure}[b]{0.28\linewidth}
    \centering
\begin{lstlisting}[language=prompt]
# Instruction
Improve the performance of the code in { target path }.

# Thought
Let's take a look at the code.

# Action
cat { target path }

# Observe
{ target snippet }

# Thought
Now let's edit the code to speed it up.

# Action
patch { target path }
@@
\end{lstlisting}
    \caption{Agentic (ReAct)}
    \label{fig:prompt:react}
\end{subfigure}%
\vspace{-0.3cm}
\caption{Templates for different prompting strategies used by \name{}.}
\end{figure*}

\niparagraph{Zero-Shot prompting.} Our prompts follow the format in Figure~\ref{fig:prompt:db:ex}. The LLM response is a code diff compatible with the {\tt patch} command. To improve performance, our fine-tuning data (Section \ref{sec:fine-tuning}) includes examples in a similar instruction-following format.

\niparagraph{Few-shot prompting.} This variant uses a small set of example edits in the prompt to guide the model, as shown in Figure~\ref{fig:prompt:fs}. We select examples of a target optimization type from our database and prompt the model to perform a similar transformation on the target code.

\niparagraph{Chain-of-Thought (CoT) prompting.} This style elicits intermediate "thoughts" from the model to trace its reasoning~\cite{wei2022chain}. By asking the model to provide ideas for improving code ``\emph{step by step}'' (Figure~\ref{fig:prompt:cot}), we guide it to identify and then implement a specific optimization strategy.

\niparagraph{ReAct prompting.} This combines reasoning with action through an iterative loop of thought, action, and observation~\cite{yao2022react} until a goal is met (Figure~\ref{fig:prompt:react}). This mimics how humans break down complex tasks. For code editing, this allows the model to examine code, propose a patch, and observe the result, providing a way to navigate ambiguity.

As we show in our evaluation, all four approaches are effective for certain tasks, but no single recipe is universally superior. We therefore maintain a library of these prompting strategies, and engineers select the most appropriate one on a case-by-case basis. Appendix~\ref{sec:supplement:llm-trajectories} provides concrete examples of these recipes and their corresponding LLM outputs.
\section{Verification, Submission, \& Monitoring}
\label{sec:verification}

Code generated by \name{} must meet production standards and achieve a high success rate to minimize the burden on human reviewers. Our edit verification process starts with automated testing. We leverage \google{}'s continuous integration pipelines, which run unit and integration tests, to check the correctness of \name{}'s edits. We automatically identify and run all tests in our entire monorepo that are affected by a change. If a change causes a build or test failure, we first attempt automated fixes for trivial errors like missing include directives. For more complex errors, we use our fine-tuned LLM to attempt a fix. If the code still fails, the edit is abandoned.

Due to imperfect test coverage, passing automated tests is not a guarantee of correctness. Furthermore, most tests do not evaluate performance and cannot detect neutral or negative efficiency impacts. We therefore incorporate LLM self-review to further improve edit quality. We prompt the model with a checklist of questions about common coding errors, anti-patterns, and optimization goals (see Appendix~\ref{sec:supplement:self-validation}). An edit proceeds to human review only after it passes this self-review step. Following \google's policies, all changes are reviewed by code owners, who make the final decision.

Management of these commits is automated using a tool called Rosie \cite{winters2020software}. Rosie allows \name{} to generate a large number of commits (called a \emph{Large-Scale Change} or \emph{LSC} in Google parlance) and then automatically identifies the correct code owners and requests reviews from them. As with any code change at Google, the commits must be reviewed and approved by code owners before they can be submitted. Google engineers are used to these kinds of reviews and perform them carefully~\cite{christopher2025instructionsetmigrationwarehouse,googleCodeReview}. This review load is accepted and generally fades into the background. \name is no different.

A range of Google tools monitor changes in production and automatically roll them back if they cause regressions~\cite{beyer2018site}. After a change is committed, we use our GWP continuous profiler (Section ~\ref{sec:background:prof}) to monitor its impact. This is non-trivial, since different projects have very different deployment strategies, and services usually roll out gradually over time. We therefore correlate the uptake of a particular commit in binaries measured by GWP with changes in metrics. We track key performance metrics to detect regressions and use monitoring alerts on performance-sensitive functions to notify us of any issues, allowing for prompt investigation. If a change correlates with a performance regression, it can be reverted. This is rare, occurring for \revertRate{} of \name's commits (Section~\ref{sec:eval}).
\section{Evaluation}
\label{sec:eval}

We evaluate \name{} through measurements from its deployment within our production fleet, and controlled experiments using examples from a small subset of our anti-patterns. We note that while we characterize \name{}'s code generation abilities, the goal is not to compare LLM generation quality across models (which has been the focus of most prior work \cite{gong2025languagemodelscodeoptimization,shypula2024learning,berger2023scalene,puri2021codenet,chen2022learning,chen2023supersonic,10.5555/3737916.3738283,liu2024evaluating,du2024mercury,pmlr-v267-huang25as}). Instead, we show how \name{} transforms a particular level of carefully characterized LLM capabilities into production impact as measured in end-to-end optimizations landing in production. As LLM capabilities are rapidly improving over time, \name{} can directly take advantages of these  improvements.

\begin{figure}[t]
\small
\lstset{aboveskip=5pt,belowskip=5pt}
  \begin{subfigure}{\linewidth}
  \centering
\begin{lstlisting}[language=code]
std::string CreateSimpleNames(
    absl::flat_hash_map<std::string, Loc> map,
    std::string key) {
  Loc value = map.at(key);
  if (value.locations.empty()) {
    return "simple_name";
  }
  return SimpleStrName(value.locations[0]);
}
\end{lstlisting}
  \vspace{-0.3cm}
  \caption{Original code snippet with an unnecessary copy anti-pattern.}
  \label{fig:eval_ubench_example:unopt}
  \end{subfigure}
  \begin{subfigure}{\linewidth}
  \centering
\begin{lstlisting}[language=code]
     std::string key) {
-  Loc value = map.at(key);
+  const Loc& value = map.at(key);
   if (value.locations.empty()) {
\end{lstlisting}
    \vspace{-0.3cm}
  \caption{Sample of a successful model generated change.}
  \label{fig:eval_ubench_example:opt}
  \end{subfigure}
  \begin{subfigure}{\linewidth}
  \centering
\begin{lstlisting}[language=code]
 std::string CreateSimpleNames(
-    absl::flat_hash_map<std::string, Loc> map,
-    std::string key) {
+    const absl::flat_hash_map<absl::string_view, 
+    Loc>& map,
+    absl::string_view key) {
   Loc value = map.at(key);
\end{lstlisting}
    \vspace{-0.3cm}
  \caption{Sample of a rejected code diff. The proposed type change requires updating the file includes but this was not added by the LLM.}
  \label{fig:eval_ubench_example:rej}
  \end{subfigure}
  \vspace{-0.3cm}
  \caption{Code optimization samples generated by \name{}.}
  \label{fig:eval_ubench_example}
\end{figure}

\begin{table*}[ht]
\caption{Microbenchmark edits. We sample the model five times for each prompt type: zero-shot (\begin{small}\texttt{ZS}\end{small}), few-shot (\begin{small}\texttt{FS}\end{small}), chain-of-thought (\begin{small}\texttt{CoT}\end{small}), and ReAct. CodeBLEU scores and speedup compare against unoptimized baselines. We report the number of avg. modified lines (\begin{small}\texttt{ModLn}\end{small}); avg. generated valid/invalid code diff hunks (\begin{small}\texttt{ValEd/InvEd}\end{small}); avg. rejected code diffs that broke code (\begin{small}\texttt{Rej}\end{small}); avg. CodeBLEU score against baselines (\begin{small}\texttt{CodeBL}\end{small}); \begin{small}\texttt{Min}\end{small}, \begin{small}\texttt{Med}\end{small}, and \begin{small}\texttt{Max}\end{small} speedups; and the speedup of the highest CodeBLEU-scoring sample (\begin{small}\texttt{CBS}\end{small}). If a generated edit is invalid, we leave the code unchanged when measuring speedup (leading to a speedup of 1). 
}
\label{table:microbenchmarks}

\newcolumntype{?}{!{\vrule width 1pt}}
\setlength\dashlinedash{0.2pt}
\setlength\dashlinegap{1.5pt}
\setlength\arrayrulewidth{0.3pt}
\small
\setlength{\tabcolsep}{2pt}
\begin{tabular}{l?l|llll:l?l|llll:l?l|llll:l|}
 \toprule
  & \textbf{Copy}  & \color{blue}ZS  & \color{blue}FS  & \color{blue}CoT  & \color{blue}ReAct & \color{teal}\color{teal}Human & \textbf{Map}  & \color{blue}ZS  & \color{blue}FS  & \color{blue}CoT  & \color{blue}ReAct & \color{teal}\color{teal}Human & \textbf{Vector} & \color{blue}ZS  & \color{blue}FS  & \color{blue}CoT  & \color{blue}ReAct & \color{teal}\color{teal}Human \\
 \midrule
\multirow{5}{*}{\rotatebox[origin=c]{90}{\textbf{Edit Metrics}}} & \begin{small}\texttt{\begin{small}\texttt{ModLn}\end{small} }\end{small}  & 2.80 & 5.60 & 15.80 & 8.00  & 6.00  & \begin{small}\texttt{\begin{small}\texttt{ModLn}\end{small} }\end{small}  & 6.60 & 7.60 & 23.60 & 9.80  & 23.00 & \begin{small}\texttt{\begin{small}\texttt{ModLn}\end{small} }\end{small}   & 2.60 & 8.60 & 63.40 & 20.40 & 3.00  \\
  & \begin{small}\texttt{ValEd}\end{small}  & 1.20 & 1.00 & 2.40  & 1.40  & 1.00  & \begin{small}\texttt{ValEd}\end{small}  & 1.00 & 1.00 & 2.80  & 1.40  & 1.00  & \begin{small}\texttt{ValEd}\end{small}   & 1.00 & 1.00 & 1.60  & 1.00  & 1.00 \\
  & \begin{small}\texttt{InvEd}\end{small}  & 0.00 & 0.20 & 0.40  & 0.00  & 0.00  & \begin{small}\texttt{InvEd}\end{small}  & 0.00 & 0.00 & 0.20  & 0.00  & 0.00  & \begin{small}\texttt{InvEd}\end{small}   & 0.00 & 0.00 & 0.20  & 0.00  & 0.00 \\
  & \begin{small}\texttt{Rej}\end{small}   & 0.00 & 0.00 & 0.40  & 0.00  & 0.00  & \begin{small}\texttt{Rej}\end{small}   & 0.00 & 0.00 & 0.60  & 0.00  & 0.00  & \begin{small}\texttt{Rej}\end{small}   & 0.00 & 0.00 & 0.00  & 0.00  & 0.00 \\
  & \begin{small}\texttt{CodBL}\end{small}  & 0.97 & 0.90 & 0.85  & 0.93  & 0.86  & \begin{small}\texttt{CodBL}\end{small}  & 0.83 & 0.83 & 0.76  & 0.84  & 0.73  & \begin{small}\texttt{CodBL}\end{small}   & 0.96 & 0.94 & 0.80  & 0.82  & 0.95 \\
  \midrule
\multirow{4}{*}{\rotatebox[origin=c]{90}{\textbf{Speedup}}}  & \begin{small}\texttt{Min}\end{small}   & 0.99 & 0.99 & 0.07  & 0.05  & 3.59  & \begin{small}\texttt{Min}\end{small}   & 1.05 & 1.05 & 1.00  & 1.04  & 1.22  & \begin{small}\texttt{Min}\end{small}   & 1.20 & 0.03 & 0.08  & 0.03  & 1.82 \\
  & \begin{small}\texttt{Med}\end{small}   & 1.01 & 1.00 & 1.00  & 1.01  & 3.59  & \begin{small}\texttt{Med}\end{small}   & 1.05 & 1.06 & 1.00  & 1.06  & 1.22  & \begin{small}\texttt{Med}\end{small}   & 1.22 & 1.21 & 0.99  & 1.00  & 1.82 \\
  & \begin{small}\texttt{Max}\end{small}   & 1.01 & 1.01 & 1.26  & 5.90  & 3.59  & \begin{small}\texttt{Max}\end{small}   & 1.07 & 1.06 & 1.07  & 1.06  & 1.22  & \begin{small}\texttt{Max}\end{small}   & 1.22 & 1.24 & 1.22  & 1.35  & 1.82 \\
  & \begin{small}\texttt{CBS}\end{small}   & 1.01 & 1.01 & 1.26  & 1.01  & 3.59  & \begin{small}\texttt{CBS}\end{small}   & 1.05 & 1.05 & 1.05  & 1.06  & 1.22  & \begin{small}\texttt{CBS}\end{small}   & 1.22 & 1.21 & 0.99  & 1.00  & 1.82 \\
\bottomrule
\end{tabular}
\end{table*}

\subsection{Microbenchmark Evaluation}
\label{sec:ubench}

We use controlled microbenchmarks to evaluate prompting strategies under conditions not feasible in production, such as testing multiple alternative edits for the same code. These experiments run on dedicated, isolated machines with many repetitions to ensure low-noise measurements. We use three hand-written C++ benchmarks, covering the \emph{Copy}, \emph{Map} and \emph{Vector} anti-patterns in Table~\ref{table:edit_categories}, and test with zero-shot, few-shot, CoT, and ReAct prompts. Each prompt asks the model to produce \emph{code diffs} to optimize the benchmark (Figure ~\ref{fig:eval_ubench_example}). For each benchmark and prompting strategy, we sample the model five times and track the number of lines modified and whether the edits are valid (i.e., the diff hunks apply cleanly).

To assess the impact of different prompting techniques, we want to obtain a diverse set of model outputs with a given prompt. However, the higher the variance of the output, the less predictable the approach. LLMs configure this trade-off through a parameter called temperature. Its value is usually between 0 and 1, where 0 is most predictable. We pick a temperature of 0.3, which we find strikes a good balance.

In all microbenchmarks, we compare against a human baseline where an expert hand-optimizes the code. Since these benchmarks are specifically written for this evaluation, none of them are in the training set of our model. One challenge in a real-world setting is that we do not know which edit performs best until we deploy it. Since it is not generally feasible to try multiple candidate edits in production, we need a way to pick a good edit prior to deploying or measuring it. This impacts the efficacy of an approach. For example, a prompting approach that occasionally provides a better optimization than other strategies does not help us if we cannot identify this optimization among its outputs. We thus need a reliable approach to pick an output to use.

To pick an output, we compare each edit with the \emph{unoptimized} baseline microbenchmarks using CodeBLEU scores which measure syntactic and semantic similarity using code-based n-gram matching~\cite{CodeBLEU2020}. The lower the score the more an edit changed the baseline's code. While this does not directly indicate the quality of an edit, it can be used to pick the output that represents the most \emph{conservative} (i.e., highest) of the generated edits, which we will see is a good strategy.

If a proposed code edit results in a benchmark that still builds, we accept the edit and mark it as valid. If it breaks the build, the edit is marked as invalid. Figure~\ref{fig:eval_ubench_example} shows examples of accepted and rejected microbenchmark code edits from \name{}. After generating edits, we apply accepted edits to the benchmarks and measure the performance in cycles per operation across 10 runs. We calculate speedup as $\texttt{speedup} = \frac{\texttt{new speed}}{\texttt{baseline speed}}$. Since edits that result in build failures are rejected and result in unchanged code, they lead to a speedup of 1. We report the minimum (\begin{small}\texttt{Min}\end{small}), median (\begin{small}\texttt{Med}\end{small}), and maximum (\begin{small}\texttt{Max}\end{small}) speedup from the five prompt samples, as well as the speedup from the most conservative edit (\emph{i.e.,} highest CodeBLEU score). Table ~\ref{table:microbenchmarks} shows the results.

The edit metrics reveal that different prompting techniques yield varying results. CoT, for example, generates the most line changes but also more invalid edits, suggesting it is better for exploration than for generating immediately submittable changes. ReAct prompting most closely mimics how programmers edit code, and achieves higher \texttt{Max} speedups, except in the case of the \begin{small}\texttt{Map}\end{small} microbenchmark. This confirms that there is no single best prompting recipe; the choice depends on the specific optimization task.

Speedup statistics also indicate that in a third of the cases, the edits achieve maximum speedups from the most conservative edits (when \begin{small}\texttt{Max}\end{small} equals \begin{small}\texttt{CBS}\end{small}, the speedup of the highest CodeBLEU-scoring
sample). In nearly all cases (except \begin{small}\texttt{Map/FS}\end{small}), \begin{small}\texttt{CBS}\end{small} meets or exceeds median speedup, highlighting that small, conservative changes can significantly improve efficiency. We thus see that conservatively picking candidates that perform a minimum amount of semantic changes to the original code selects edits that lead to good efficiency improvements. While humans still outperform the AI, superhuman capability is not required for \name{}, since the key benefit of AI is its ability to cover much more code.

\begin{table}[t]
\caption{Comparison of Bag of Words (BOW), Deep Text Embedding (DTE), and Deep Code Embedding (DCE), with and without ranking, for identifying optimization opportunities.}
\label{table:retrieval}
\begin{center}
\small
\scalebox{0.9}{
\begin{tabular}{ c|cc|ccc } 
\toprule
\textbf{Model} & \textbf{Query} & \textbf{Ranked} & \textbf{\begin{small}\texttt{MAP@5}\end{small}} & \textbf{\begin{small}\texttt{MAP@10}\end{small}} & \textbf{\begin{small}\texttt{MAP@20}\end{small}} \\
\midrule
BOW & Function & No & 0.0287 & 0.0173 & 0.0128 \\
BOW & Function & Yes & 0.0728  & 0.0398 & 0.0224 \\
\midrule
DTE & Function & No & 0.0259 & 0.0190 & 0.0137 \\
DTE & Function & Yes & 0.1633 & 0.0947 & 0.0554 \\
\midrule
DCE & Function & No & 0.1945 & 0.1585 & 0.1111 \\
DCE & Function & Yes & 0.2036 & 0.1316 & 0.0918 \\
\bottomrule
\end{tabular}
}
\end{center}
\end{table}

\subsection{Candidate Retrieval}
\label{sec:embed}

We now evaluate the effectiveness of \name's candidate retrieval methods (Section~\ref{sec:localization:code-sim}). We create a test set comprising 63 pairs of code diffs and corresponding ``\emph{before}'' functions. This set includes 21 pairs each for unnecessary copies, redundant map lookups, and missing vector reserves.

To assess retrieval performance, we build a database of 1803 functions, including functions from the 63 test pairs and an additional 1740 functions without known opportunities for the targeted optimizations. For each of the 63 test functions, we use its code embedding to query the database. A successful retrieval model would rank functions of the same anti-pattern higher when queried with a function of that anti-pattern. We compare the retrieval models in terms of mean average precision (\begin{small}\texttt{MAP}\end{small}), a standard measure for evaluating retrieval models ~\cite{schutze2008introduction}. A query and a function are deemed ``\emph{relevant}'' if they share the same optimization opportunity.

\begin{table}
\caption{Prompting strategy edit metrics. Modified lines (ModLn), valid/invalid/exact edits (ValEd/InvEd/ExtEd), and CodeBLEU (CodBL) and quality scores (QualS) are shown as averages per prompting technique. 
}
\label{table:eval_edit_gen}
\scalebox{0.9}{
\begin{tabular}{ c|cccccc } 
\toprule
\textbf{Prompt} &\textbf{\begin{small}\texttt{ModLn}\end{small}} & \textbf{\begin{small}\texttt{ValEd}\end{small}} & \textbf{\begin{small}\texttt{InvEd}\end{small}}& \textbf{\begin{small}\texttt{ExtEd}\end{small}} & \textbf{\begin{small}\texttt{CodBL}\end{small}} & \textbf{\begin{small}\texttt{QualS}\end{small}} \\
\midrule
ZS & 9.392          & 1.752       & 0.128         & 0.428          & 0.967 & 0.531           \\
FS & 12.736          & 1.796       & 0.156         & 0.088        & 0.958  & 0.404           \\
CoT & 22.644          & 3.272       & 0.82          & 0.152        & 0.954   & 0.344           \\
ReAct & 10.444          & 1.804       & 0.1           & 0.264        & 0.969 & 0.510  \\         
\bottomrule
\end{tabular}
}
\end{table}

\begin{figure*}[t]
\centering
\includegraphics[width=\linewidth]{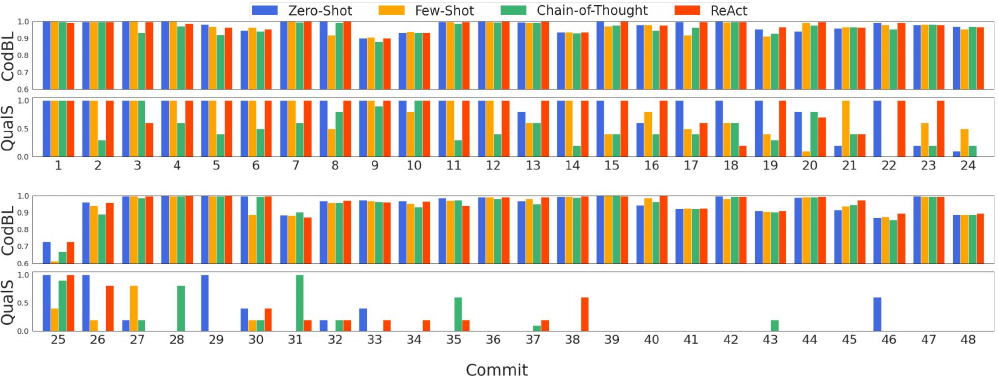}
\label{fig:overview:example}
\vspace{-15pt}
\caption{Quality of generated edits by prompting strategy.
CodeBLEU scores (CodBL) show comparison to true historical changes. Quality scores (QualS) show code edit quality as scored by human experts reviewing whether the code edit is correct and performance-improving (Rubric: 1 = Yes, 0.5 = Partially, 0 = No). In total, human raters scored 960 examples.}
\label{fig:eval_edit_gen}
\end{figure*}

\begin{table*}[t]
\caption{Feedback from human and automated code-review.}
\label{table:casestudies}
\scalebox{1}{
\setlength\dashlinedash{0.2pt}
\setlength\dashlinegap{1.5pt}
\setlength\arrayrulewidth{0.3pt}
\footnotesize
\begin{tabular}{@{}l|ccc|p{0.53\linewidth}
} 
\toprule
 & \textbf{Copy}  & \textbf{Map} & \textbf{Vector} & \textbf{Description}
                    \\
\midrule
\textbf{S\_PROD}         & 39.97\%   & 4.86\%        & 40.89\% &Submitted to production.  
\\\hdashline
\textbf{S\_USER}   & 6.55\%    & 10.49\%       & 15.99\% &Submitted with user discussion (e.g., reviewer questions or suggested changes).
\\\hdashline
\textbf{R\_REVERT} & 0.16\%    & 0.14\%        & 0.12\%  &Submitted to production but later rolled back due to regressions.    
\\\hdashline
\textbf{R\_TEST}   & 16.14\%   & 37.39\%       & 20.07\%  &Rejected during the validation phase.    
\\\hdashline
\textbf{R\_USER}   & 35.81\%   & 46.55\%       & 22.87\% &Rejected by human reviewers.     
\\\hdashline
\textbf{R\_EMPTY}  & 0.00\%    & 0.07\%        & 0.00\%  &Rejected due to failure to make code modifications beyond formatting.    
\\\hdashline
\textbf{R\_OTHER}  & 1.37\%    & 0.49\%        & 0.05\%  &Rejected for other reasons, typically non-edit related, such as failure to resolve merge conflicts during submission or failure to identify proper reviewers.   
\\
\midrule
\textbf{Total Commits}     & 4959      & 1421          & 4035  & Total number of code commits that were generated.    
\\\hdashline
\textbf{NC Savings}    &   331k         &       125k    &    618k & Efficiency impact of the changes in normalized cores.   
\\
\bottomrule
\end{tabular}
}
\end{table*}

Table~\ref{table:retrieval} presents the performance of these retrieval models using different query types. \begin{small}\texttt{MAP@K}\end{small} quantifies retrieval quality by considering the top-$\mathcal{K}$ results for each query. When $K=5$, bag of words (BOW) only achieves low \begin{small}\texttt{MAP@K}\end{small} scores even after functions are ranked by syntactic similarity scores (Section ~\ref{sec:localization:code-sim-ranking}). The MAP scores drop even further when $K$ increases to 10 and 20. In contrast, \begin{small}\texttt{MAP@5}\end{small} more than doubles from 0.0728 to 0.1633 with function rankings on the deep text embedding model (DTE) and 0.2036 with the deep code embedding model (DCE).

Deep text embeddings show significant improvements over BOW, despite their lack of explicit code training. This suggests that text embedding models inherently capture some code semantics. Deep code embeddings show further improvements over deep text embeddings, which is expected from models trained on semantically similar function pairs.

Finally, applying a second-step syntactic ranking boosts performance for the BOW and text embedding models, underscoring the value of syntactic comparison when a model lacks full code understanding. The deep code model, however, shows little benefit from this re-ranking, suggesting that its end-to-end training already captures the necessary semantic and syntactic information. These findings validate the effectiveness of our embedding-based approach for retrieving optimization candidates.

These findings show the efficacy of \name{}'s embedding-based approach to code optimization candidate retrieval.

\subsection{Edit Generation on Production Code}
\label{sec:edit_gen}

While microbenchmarks allow easy measurement of performance improvements, they do not capture the complexity of a large-scale production code base. We now assess the quality of generated edits on production code. We focus on the four prompting strategies applied to 48 performance-improving commits from our real code repository (not used to fine-tune our LLM). We aim to determine how accurately each prompt replicates the chosen code edits.

Contrary to the microbenchmark study, we cannot perform isolated performance tests on each generated edit. We thus evaluate these edits along multiple dimensions, each of which captures a different aspect of correctness.
First, \emph{validity} captures whether the code compiles (i.e., is syntactically correct).
Second, \emph{exact match} to the original edit is a very strong signal for correctness when there is a match, but provides little signal otherwise.  Third, \emph{CodeBLEU} scoring with respect to the original edit provides a signal how close the edit is syntactically, but small syntactic changes can make a large difference for correctness (e.g., including an additional \verb|&| or \verb|*|). To capture these subtleties, we also obtain a \emph{quality score} for each edit through human scoring. While work-intensive, the human scoring is critical to provide ground truth, as automated scoring strategies can only provide limited insights in this evaluation setting. This is in line with best practices \cite{DBLP:journals/corr/abs-2002-08512}.

Given a target performance-improving commit, we use zero-shot, few-shot, CoT, and ReAct prompts to provide the model with the original source code and instructions to optimize the code. For each prompt, we sample our model five times. Each generated edit is then evaluated using CodeBLEU scores and by humans.
Figure~\ref{fig:eval_edit_gen} breaks down the CodeBLEU and human-reviewed code quality scores for each commit, while Table~\ref{table:eval_edit_gen} shows averages across the generated edits. 

Zero-shot and ReAct prompting achieve the highest CodeBLEU and human-scored quality scores. In contrast, CoT provides the lowest CodeBLEU and quality scores. Figure~\ref{fig:eval_edit_gen} indicates that CodeBLEU scores generally align with human scores, trending higher with better code quality scores. However, discrepancies arise when the model proposes minor, non-performance-improving changes (\emph{e.g.}, altering variable names or formatting \textemdash{} commits 47 and 48) in which the edit receives a high CodeBLEU score but a low quality score. On the other hand, if the model generates performance-improving changes but also adds extensive documentation (\emph{e.g.}, commit 25), the edit receives a low CodeBLEU score but high quality score. These results suggest that CodeBLEU can provide a weak signal for performance quality when comparing against known performance-improving edits. This study also reveals that insights from microbenchmarks hold true for production edits. CoT prompts produce larger edits but do not always result in the highest quality while targeted approaches like ReAct prompting tend to be more accurate.

\subsection{Production Impact Evaluation}
\label{sec:lifecycle}

We now analyze \name{}'s production impact and its interactions with Google's infrastructure and processes. \name{} has generated thousands of edits to date. We measure compute savings in \emph{normalized cores} (NC), defined as the MIPS provided by a single core on a specific hardware platform~\cite{borg2020}. In production, our approach employs a mix of zero-shot, few-shot, CoT, and ReAct prompting strategies on a case-by-case basis. We focus once again on the \begin{small}\texttt{Copy}\end{small}, \begin{small}\texttt{Map}\end{small} and \begin{small}\texttt{Vector}\end{small} anti-patterns.

\niparagraph{Edit Quality} We quantify the quality of \name{}-generated edits at scale by analyzing their code reviews and deployment, which includes human feedback, reverts due to production issues or incorrect changes, and resultant performance improvements using our fleet-wide performance profiler. For every change generated by \name{}, a human code owner is asked to review the change before the change is submitted to production. For 40\%, 5\%, and 41\% of \begin{small}\texttt{Copy}\end{small}, \begin{small}\texttt{Map}\end{small} and \begin{small}\texttt{Vector}\end{small} changes, respectively, \name{}-generated changes were directly approved and submitted to production without the human reviewers leaving any feedback that had to be resolved. 

Table~\ref{table:casestudies} shows a breakdown of additional actions required (see Appendix~\ref{sec:supplement:experirence-reports} for concrete reviewer interactions). Oftentimes, users left comments that were extraneous to the optimization in the code change (\verb|S_USER|). These comments had to be addressed before the code change was submitted. Other times, users rejected changes (\verb|R_USER|) and the changes were not submitted. This could be due to the reviewer judging the change to not be performance enhancing or wanting the code to be untouched for other reasons (e.g., style preferences).

\niparagraph{Failure Modes} The reasons why optimizations failed were multi-faceted. For example, \texttt{Map} has a comparatively low success rate. While \texttt{Copy} and \texttt{Vector} changes are relatively localized, \texttt{Map} changes typically require restructuring and reasoning about large pieces of code while also dealing with subtle semantics of C++ maps. Examples of errors include:
\begin{itemize}
 \item \textbf{Control Flow}: Rewrites often caused subtle changes in behavior along some execution paths (e.g., adding an element too early or too late, confusing the behavior of the [] operator vs. \texttt{try\_emplace}, \texttt{contains}, etc.).
 \item \textbf{Pointer Stability}: The model sometimes assumed that a pointer to a map entry remains valid throughout map modifications even if a \texttt{flat\_hash\_map} is used.
 \item \textbf{Corner Cases}: E.g., \texttt{if (!a.contains(k)) \{ a[k] = v \} else \{ a[k] = min(v, a[k]) \}} and \texttt{a[k] = min(v, a[k])} are not equivalent.
 \item \textbf{Not a Map}: Sometimes the model thought that something that was not a map was a map (e.g., a proto).
 \item \textbf{No Change}: The change was semantically equivalent or did not improve performance (e.g., replacing a single [] with a \texttt{try\_emplace}).
\end{itemize}

\noindent Our automated pre-screening filters a significant portion of these before they reach human reviewers, reducing the review burden. Sometimes, we found that it was the reviewer who was mistaken. For example, we saw a case where the model added a \texttt{v.reserve(2)}, causing the reviewer to reject the change as erroneous. However, later in the code, there was an assertion \texttt{ASSERT\_EQ(2, v.size())}, indicating that the change, while unexpected, would have been correct. In one case, a correct \name{} code change optimizing a redundant map lookup was sent to a reviewer who suggested using \texttt{try\_emplace} instead (Appendix~\ref{sec:supplement:experirence-reports}). The reviewer's suggestion was applied instead of the original \name{} change. This led to large increases in CPU utilization and had to be reverted (\texttt{R\_REVERT}). This highlights the subtleties of human-AI collaboration.

\begin{figure}[t]
    \centering
    \begin{subfigure}[t]{1\linewidth}
        \centering
        \includegraphics[width=1\linewidth]{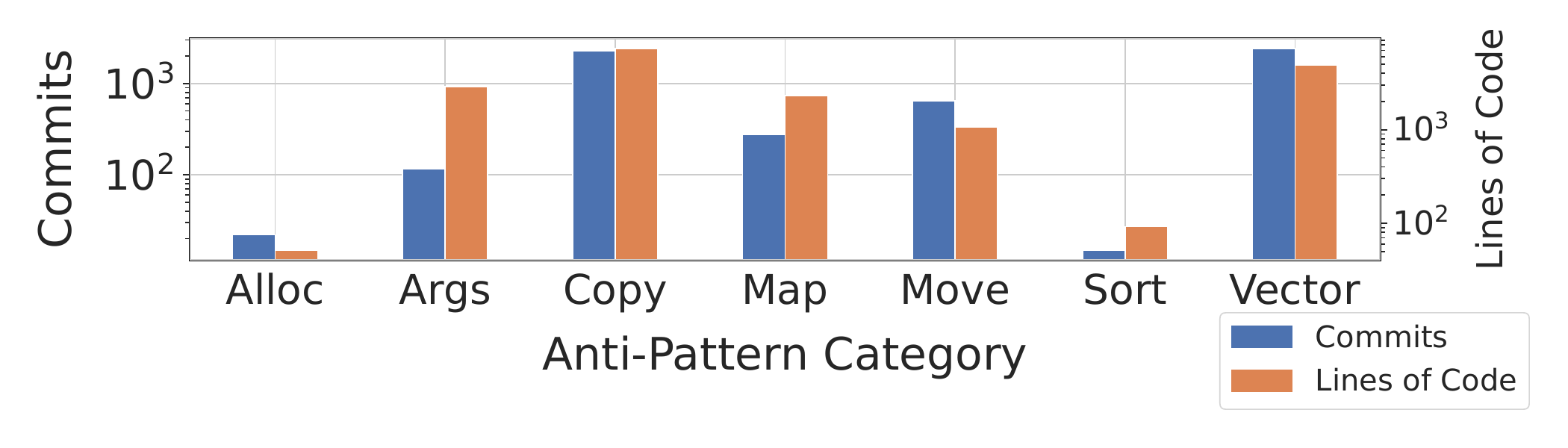}
    \end{subfigure}
    \vspace{-0.2cm}
    \begin{subfigure}[t]{1\linewidth}
        \centering
        \includegraphics[width=\linewidth]{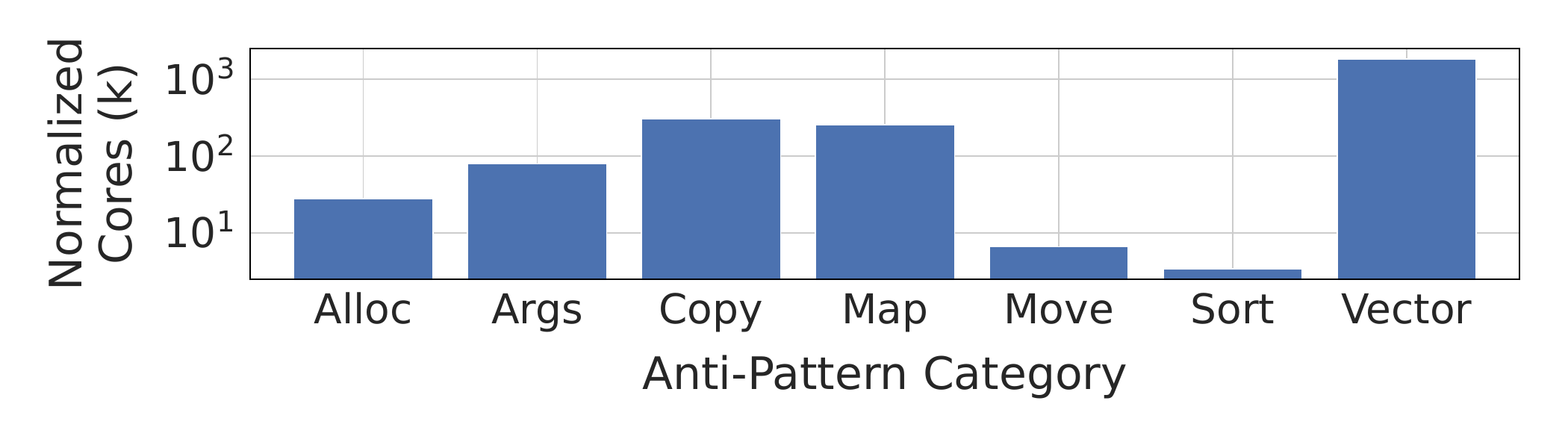}
    \end{subfigure}
    \vspace{-0.2cm}
    \begin{subfigure}{\linewidth}
        \centering
        \includegraphics[width=1\linewidth]{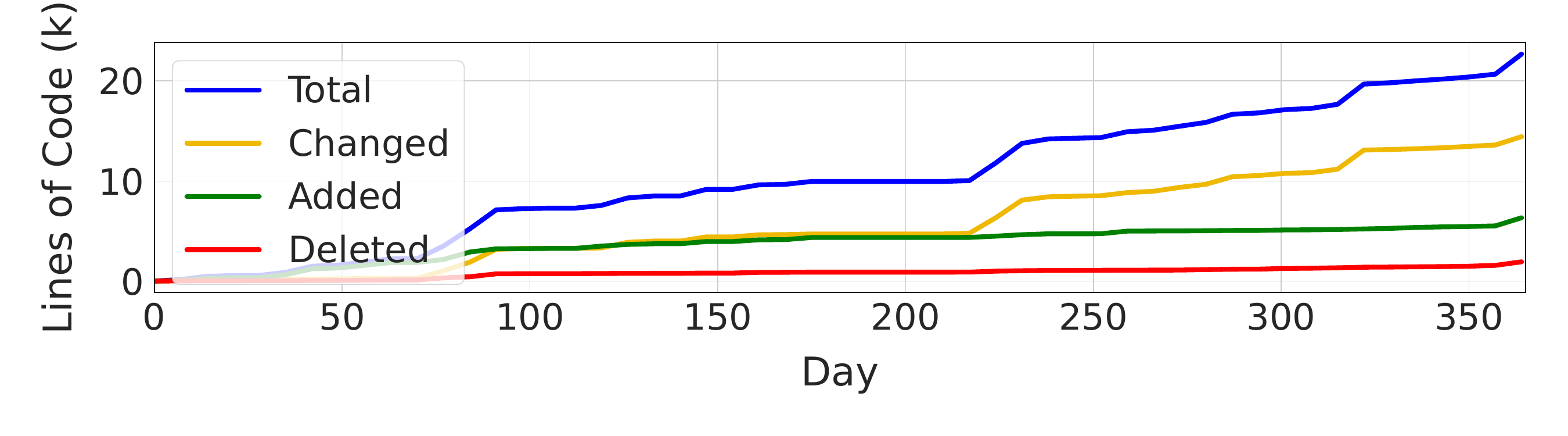}
    \end{subfigure}
    \caption{Statistics on landed code edits. Top: submitted code edits. Middle: performance savings. Bottom: landed code edits in production over time.}
    \label{fig:eval:case_study}
\end{figure}

\niparagraph{Resource Savings} Figure~\ref{fig:eval:case_study} shows the savings achieved. Savings per line of code also vary by anti-pattern. Complex \begin{small}\texttt{Map}\end{small} changes require more lines of code and yield lower savings per line than simple \begin{small}\texttt{Copy}\end{small} or \begin{small}\texttt{Vector}\end{small} changes. In contrast, \begin{small}\texttt{Vector}\end{small} edits are often single-line changes but have a high impact, as they can prevent many reallocations of a large data structure.

Over a period of one year, \name{} has sent commits covering many anti-patterns (Section~\ref{sec:anti-patterns}), worth over several hundred thousand normalized CPU cores.  Figure~\ref{fig:eval:case_study} shows the volume of code changes \emph{successfully} submitted, with a rollback rate of less than 0.5$\%$. In total, \name{} has submitted over \totalLinesOfCodeChanged lines of code across \totalCommits commits. The majority of these changes modify or add lines, and some edits also remove lines. These results demonstrate the potential of \name{} to achieve fleet-wide performance improvements with minimal human effort.

\niparagraph{Cost Analysis} An important question is whether the benefits of \name{} justify its operational costs. Conservatively using prices for Gemini 3.1 Pro Preview, the most expensive Google model at the time of writing (\$2-12 per 1M tokens \cite{gemini-cost}) and conservatively assuming 10k tokens (90\% input) per invocation and 10 invocations per commit (including self-review and refinement), generating 10k commits costs \$3,000.

Engineering and infrastructure costs for code review are not unique to \name{} and similar to the background review load generated by other automated tools \cite{christopher2025instructionsetmigrationwarehouse}, which is generally accepted at Google. Overall, the additional human and infrastructure load for submitting changes from \name{} was negligible. While we cannot share the exact number, it represented less than 0.1\% of overall resources. This is a one-off cost, while the resource savings are realized on an ongoing basis.

\begin{table*}[ht]
\centering
\caption{Static Analysis vs. \name{} vector reserve optimizations. Speedup ranges represent empirical CPU time reductions from micro-benchmarks (* denotes qualitative analysis where no runnable artifact was available).}
\label{tab:comparison_clang_tidy_eco}
\makebox[\textwidth][c]{%
\resizebox{\textwidth}{!}{%
\setlength\dashlinedash{0.2pt}%
\setlength\dashlinegap{1pt}%
\setlength\arrayrulewidth{0.3pt}%
\setlength{\aboverulesep}{0pt}%
\setlength{\belowrulesep}{0pt}%
\renewcommand{\arraystretch}{1.4}%
\setlength{\tabcolsep}{8pt}%
\footnotesize
\begin{tabular}{
    @{} 
    >{\raggedright\arraybackslash}m{2.1cm}
    @{\hspace{6pt}\vrule width 0.4pt\hspace{6pt}} 
    >{\centering\arraybackslash}m{1cm} 
    >{\centering\arraybackslash}m{1cm} 
    >{\centering\arraybackslash}m{1.2cm} 
    >{\centering\arraybackslash}m{1.2cm} 
    >{\raggedright\arraybackslash}m{1.8cm} 
    >{\scriptsize}m{4cm} 
    >{\raggedright\arraybackslash\scriptsize\ttfamily}m{5.3cm}
    @{} 
    }
\toprule
\normalfont\textbf{Pattern} & \textbf{ClangTidy} & \textbf{CppCheck} & \textbf{CARAMEL*} & \textbf{CLARITY*} & \textbf{\name Speedup} & \normalfont\footnotesize\textbf{Explanation} & \normalfont\footnotesize\textbf{Example Code Snippet} \\
\midrule

\textbf{A: Standard Loop Insertion} & 
Yes & No & Yes & Yes & 0.4\% -- 7.4\% & 
Trivial loop structure matching directly on standard library vector insertion (\texttt{.push\_back}). &
dirs.reserve(src.size());\newline for (auto g : src) \newline \hspace*{1em}dirs.push\_back(g); \\
\midrule

\textbf{B: Bulk Move via Algorithms} & 
No & No & No & No & 2.1\% -- 13.8\% & 
Elements are appended via algorithms/wrappers without an explicit local loop in the AST. &
res.reserve(res.size() + src.size());\newline absl::c\_move(src,\newline \hspace*{1em}std::back\_inserter(res)); \\
\midrule

\textbf{C: Custom Container Assignment} & 
No & No & No & No & 5.6\% -- 9.9\% & 
Target is a protocol buffer; insertion bypasses standard vector analysis via implicit allocations in assignment operators. &
// RepeatedPtrField assignment\newline cache.mutable\_ents()->Reserve(size);\newline *cache.mutable\_ents() = \newline \hspace*{1em}\{src.begin(), src.end()\}; \\
\midrule

\textbf{D: Template Range Copies} & 
No & No & With LTO & Yes & 1.4\% -- 4.1\% & 
Container types are generic template arguments unresolved at static analysis compile-time. &
template <class R> void Copy(const R\& s) \{\newline \hspace*{1em}std::vector<Ent> d;\newline \hspace*{1em}d.reserve(s.size());\newline \hspace*{1em}d.insert(d.end(), s.begin(), s.end());\} \\
\midrule

\textbf{E: Inter-Procedural Size Maps} & 
No & No & Partial & Partial & 3.0\% -- 13.7\% & 
Sizing context is decoupled. Conditional insertions prevent mapping to \texttt{src.size()}; a pre-calculated stat dictates the true needed capacity. &
// st.total\_size == count of ok items\newline d.reserve(st.total\_size);\newline for (const auto\& x : src)\newline \hspace*{1em}if (x.ok) d.push\_back(x); \\
\midrule

\textbf{F: Custom Pointer Array Buffers} & 
No & No & No & No & 3.2\% -- 12.7\% & 
Target is a custom API class. Static analysis lacks semantic awareness that \texttt{.Add()} stores pointers and triggers underlying array expansions. &
CustomPtrBuffer buf;\newline buf.Reserve(src.size());\newline for (auto\& x : src) \newline \hspace*{1em}buf.Add(\&x); \\
\bottomrule

\end{tabular}%
}%
}
\end{table*}

\section{Related Work}
\label{sec:related}

\niparagraph{Code optimization} Automatic code optimization has a long history. Stochastic superoptimization~\cite{schkufza2013stochastic} explores instruction sequences to find optimal solutions, operating below the source code level. More recently, AlphaTensor~\cite{fawzi2022discovering} and AlphaDev~\cite{mankowitz2023faster} use reinforcement learning for algorithm discovery for specific problems. The Scalene profiler~\cite{berger2023scalene} interfaces with ChatGPT to optimize Python code. PIE~\cite{shypula2024learning} extends the CodeNet Benchmark~\cite{puri2021codenet} and designs evaluation methodologies and prompting techniques for performance improvements. Chen et al.~\cite{chen2022learning} used VAE-based models on Google Code Jam data, while Supersonic~\cite{chen2023supersonic} applies Seq2Seq models to Codeforces~\cite{codeforces} and the CodeNet dataset~\cite{puri2021codenet}.

Some works attempt to scale these methods beyond competition datasets. Aroma~\cite{luan2019aroma} provides general code recommendations across a code base but does not focus on performance. Garg et al.~\cite{garg2022deepdev} train models on C\# performance-improving edits and evaluate on open-source repositories, but focus less on scalability and real-world deployments. Other work like EditLord~\cite{editlord_icml25} also fine-tunes LLMs on code edits, but this improves an LLM's capability at code editing as opposed to solving localization and reliability. To the best of our knowledge, \name{} is the first to demonstrate and evaluate code optimizations at scale, applying them to \linesofcode{}. Unlike prior work, we emphasize automation and the warehouse-scale setting, resulting in thousands of real-world performance improvements.

\niparagraph{Code transformations} Mining and applying code transformations in existing code is not a new problem. Meng et al.~\cite{meng2011systematicedits} build a sophisticated custom algorithm to generalize from an example and then apply the ``same'' transformation to other locations. More recently, Miltner et al.~\cite{miltner2019onthefly} present an algorithm to synthesize a script from repetitive changes and enable automated propagation of those changes. In our setting, examples of changes are already provided as anti-patterns, and we find that modern LLMs are capable enough to apply these changes at other locations, with no need for custom algorithms. Overall correctness must be verified, but this is also true of these prior AST-based techniques. Finally, Uber's DR.FIX system \cite{10.1145/3729265} also uses LLMs for edits, but applies them to data races in Go, a very different problem.

\niparagraph{Repository-level refactoring} Our work also relates to recent work in rolling out repository-level code changes, such as CodePlan \cite{bairi2024codeplan}. While superficially similar, the primary commonality is that the actual transformation is carried out by LLMs. However, the emphasis of CodePlan is to apply a precisely specified edit that requires modifications to multiple files, such as changing a function signature or type, leveraging static analysis to identify additional locations that need to be changed. In contrast, our work focuses on \emph{identifying} opportunities for code optimization across a repository of \linesofcode{}, and \emph{generalizing} existing edits to these new contexts. Our work is carried out at a much larger scale---CodePlan's largest repository had $\approx$ 20k\xspace lines of code.

\section{Discussion \& Future Work}
\label{sec:discussion}

\niparagraph{Deployed SOTA} \google~relies on compilers for optimization, which cannot apply semantics-altering changes like \name{}. We also deploy LLVM's Clang-Tidy~\cite{clangtidy}, which uses heuristics to detect errors, including performance issues. However, we found it has notable limitations.

\niparagraph{Static Analysis} We compare \name{} to static analysis tools. Table~\ref{tab:comparison_clang_tidy_eco} summarizes vector reserve patterns—a primary anti-pattern category for both static tools and \name{}—and whether tools can optimize them. We include Clang-Tidy~\cite{clangtidy} and CppCheck~\cite{cppcheck}, two widely deployed static analysis tools. We also perform a qualitative comparison to CARAMEL~\cite{caramel2015} and CLARITY~\cite{clarity2015}, two published static analysis techniques that target performance optimizations (but have no artifacts).

We group common code sequences that can be optimized with vector reserve operations into 6 patterns. For each pattern, we run microbenchmarks spanning 8 to 4,096 container elements and report the resulting speedups, confirming that these changes represent real and significant optimization opportunities. Out of the 6 patterns, Clang-Tidy reliably optimizes only 1, and CppCheck optimizes none, whereas \name{} successfully optimizes all 6. The reason is that the first pattern is explicitly encoded in Clang-Tidy as the standard ``\textit{performance-inefficient-vector-operation}''~\cite{clangtidy-vector} check, which identifies simple patterns such as \begin{small}\verb|for (auto e : data) v.push_back(e);|\end{small}.
In those cases, Clang-Tidy adds a call to \begin{small}\verb|v.reserve|\end{small}, but this only works if the control flow and structure are trivial.\footnote{Clang-Tidy documentation states that ``\textit{the check only detects [...] loops with a single statement body}''~\cite{clangtidy-vector}. If the loop is more complex (e.g., a while loop or function calls), the pattern matching fails.}

To determine if this is a shortcoming of specific tools or a fundamental static analysis limitation, we examine CARAMEL~\cite{caramel2015} and CLARITY~\cite{clarity2015}, two static analysis tools that specifically target loop-related performance optimizations. Directly applying either tool would find none of the 6 patterns, since they are not the target of their analyses. CARAMEL~\cite{caramel2015} focuses on wasted loop iterations and CLARITY~\cite{clarity2015} identifies redundant traversal bugs, neither of which are present in the examples in Table~\ref{tab:comparison_clang_tidy_eco}. We hence only focus on the key ideas of these papers and try to adapt them.

Patterns B, C, and F are out of reach for static analysis since they rely on external functions and domain specific knowledge that are not available to the static analyzer. Pattern D requires templates to have been fully resolved by the time static analysis is applied; they may be within reach with LTO \cite{7863733}. Pattern E may be within reach for either approach if the \texttt{x.ok} field matches one of the patterns it is searching for.

This points to a fundamental limitation of all static analysis tools – they can only identify the specific optimizations that they were designed for. Just like Clang-Tidy and CppCheck, CLARITY and CARAMEL only target specific optimizations. However, within this context, they have a key advantage: It is easier for them to perform inter-procedural analysis that spans disparate locations in the code, a task that used to be challenging for LLMs. Recent work proposed combining static analysis and LLMs to achieve the best of both worlds \cite{10.1145/3652588.3663317}.

Expanding static analysis via custom rules for all corner cases requires prohibitive human engineering effort, quickly surpassing potential resource savings. CodeQL \cite{youn2023declarative} facilitates this but does not solve it. In contrast, \name{} is designed to expand beyond a specific optimization and can perform diverse optimizations via examples, eliminating the need to implement new analysis passes for each optimization.

\niparagraph{Deployment Insights} One of the early challenges for \name{} was to build trust with engineers. We thus prioritized our validation pipelines early on and picked initial edits carefully.

\name{}'s production success rate exceeded initial expectations, and reviewing changes from \name{} helped shift some of our perspectives on the viability of optimizing code at Google with LLMs. Historically, LSCs were generated using deterministic tools, which restricted them to rigid, highly repetitive patterns. \name{} helped us expand what we could optimize with automation. However, the increased diversity of commits meant that the team had to carefully monitor and iterate on the review feedback. This feedback loop was critical.

\name{} depends heavily on the capabilities of the underlying model. Early on, we developed an evaluation to track the model capabilities on a set of canonical commits and understand what capabilities we could expect. We found it important to expand capabilities incrementally. Some optimization patterns were out of reach initially but later became feasible. E.g., optimizations such as proto arenas (Appendix~\ref{sec:supplement:multi-file-edits}) require modifications across many files, which was not possible in early versions of \name{} but became feasible over time.

While there are many prompting strategies, we found no clear rule on which works best when. Selecting strategies was highly empirical and sometimes combined techniques.

Some anti-pattern categories are much harder than they appear, e.g., \texttt{Sort}. Determining whether a sorted map or set can be replaced with an unsorted version requires judgement and significant context that early LLMs struggled at.

The embedding-based retrieval approach, when scaled to the scale of Google's multi-billion line code base, required careful calibration. There is a trade-off between precision and recall, and getting this balance wrong risks wasting significant reviewer time or missing important opportunities. We found that the most problematic candidates are those that seem plausible but are false positives.

As \name{} has been successfully optimizing simple inefficiencies, the remaining opportunities have shifted toward more complex refactorings. These targets yield higher false-positive rates during candidate retrieval (e.g., heap-allocated protobufs that are candidates for arena migration can look semantically similar to code that is not, such as when an external pointer lifetime prohibits safe migration). To catch these false positives, we have begun deploying agent-based LLM validators during opportunity localization.

\niparagraph{Future Work} Our approach has several potential extensions. Currently, we use a \emph{top-down} approach, leveraging an anti-pattern database to find optimization candidates. A complementary \emph{bottom-up} approach could start from profiling data and then use an LLM to diagnose and fix bottlenecks. Further, our current \emph{reactive} approach improves performance post-deployment. A similar framework could be used \emph{proactively} to suggest edits to developers as they write code.

We believe \name{}'s core techniques can be replicated elsewhere using openly available tools and LLMs (Appendix~\ref{sec:supplement:reproducibility}) and apply to smaller and decentralized codebases. We think the primary challenges to wider adoption are often organizational rather than technical, requiring integration with existing developer workflows and review processes.
\section{Conclusion}
\label{sec:conclusion}

We described \name{}, a system for automatic code optimization at data center scale. By assembling a dataset of over \numberOfAntipatternsInTraining{} performance-improving edits from our code repository, we use code similarity search to identify similar patterns, and LLMs to apply changes automatically. Deployed in production at Google, \name{} has changed more than \totalLinesOfCodeChanged lines of code, saving hundreds of thousands of normalized CPU cores.

\section*{Acknowledgements}
\label{sec:acknowledgements}

We would like to especially thank Chandu Thekkath, David Lo, Fredrik Kjolstad, James Laudon, Liqun Cheng, Mangpo Phothilimthana, Niranjan Tulpule, Steve Blackburn, Tipp Moseley and the anonymous reviewers for their feedback. We would also like thank our shepherd for their help in improving the final version of the paper.

\clearpage
\pagebreak
\appendix

\onecolumn
\section*{Supplementary Material}

In this appendix, we provide additional details to the main paper, expanding on key areas such as a comprehensive list of anti-patterns identified, full LLM trajectory examples showcasing various querying approaches, multi-file edit considerations, LLM self-validation examples, human code reviewer interactions, and a discussion on the reproducibility of this work. This appendix is provided to offer a deeper understanding of the methodology, challenges, and outcomes of \name{}'s approach to automating code efficiency optimization.

\section{Anti-Patterns}
\label{sec:supplement:anti-patterns}

\begin{figure}[htbp]
    \centering
    \includegraphics[width=\linewidth]{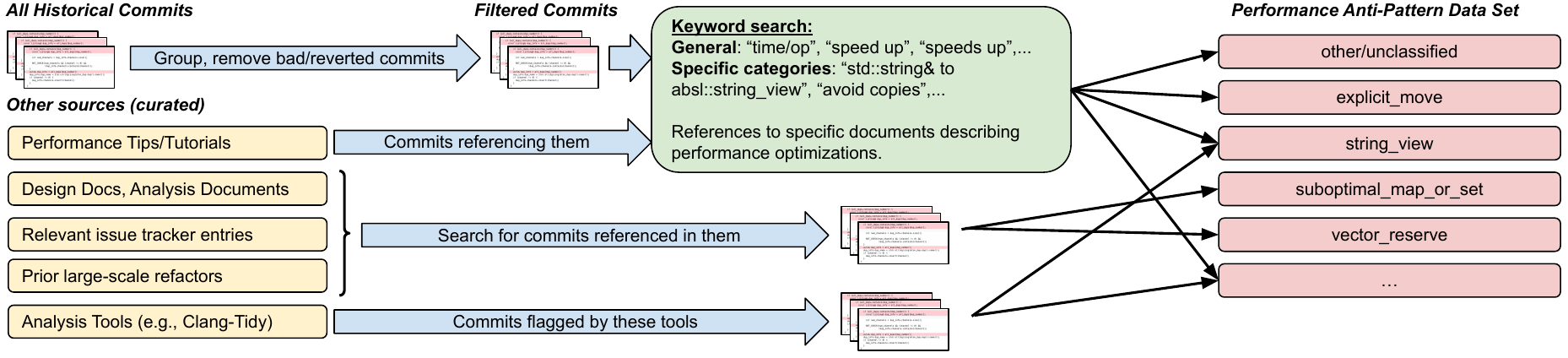}
    \caption{The approach and processing pipeline used to mine performance anti-patterns from our repository's commit history and a number of curated additional data sources.}
    \label{fig:anti-patterns:pipeline}
\end{figure}

Figure~\ref{fig:anti-patterns:pipeline} summarizes how \name{} takes historical code commits and other code sources and turns them into an anti-patterns dataset (Section~\ref{sec:anti-patterns}). This provides additional context and examples for the description in the main paper.

We also provide a full list of anti-pattern categories extracted in the dataset in Table~\ref{table:supplemental-anti-patterns}. Note that in the main paper, we provide more details on a subset of these categories, but include the full table here for completeness. As new anti-patterns are discovered, the table can grow over time.

\clearpage

\begin{table}[htbp]
\captionof{table}{List of all
Anti-Pattern Categories}
\label{table:supplemental-anti-patterns}
\footnotesize
\begin{tabular}{p{0.2\linewidth}| p{0.75\linewidth} }
\toprule
\textbf{Category} & \textbf{Description} \\
\midrule
Atomics & Use of std::atomic -- std::atomic should be only used for simple properties where the value is returned as is and the property is set at most once with a mutex lock. \\
\hline
Callables & Unnecessary copies of a callable. \\
\hline
Data counters & Inefficient use of data processing counters. \\
\hline
Expensive Calls & Return early, cache results, or change the calling frequency of an expensive function call. \\
\hline
Futures & Using future allows users to treat future-producing APIs as cheaply copyable values with value semantics. \\
\hline
Hash Tuning & Improving hash tables with too many collisions or bits that never change across every hash with well-mixed hash functions provided by Abseil (\url{https://github.com/abseil/abseil-cpp}). \\
\hline
Hashes & Use of an inefficient hashing library when a more efficient option exists. \\
\hline
Heterogeneous Lookups & Map insertions/lookups causing the conversion of near-equivalent types due to user-specified comparator function to the map. \\
\hline
Intrusive Collection & Use elements of the collection to store state related to the implementation of the collection. \\
\hline
Lambda Capture & Capture by moving rather than copying and capturing individual member variables of an object. \\
\hline
Locks & Use of an inefficient lock type. \\
\hline
Missing  \verb|std::move|s & The last use of an object is copied instead of moved. \\
\hline
Missing Vector Reserves & See Figure~\ref{fig:dataset:anti-patterns:c}. \\
\hline
Optional & Use the std::optional when an object can be allocated on the stack but its population is deferred subsequent to construction. \\
\hline
PRNGs & Pseudorandom number generators that can use more efficient random number generation libraries. \\
\hline
Proto Arenas & Reduce the allocation and deallocation costs of protocol buffers by allocating them on a large piece of pre-allocated memory. \\
\hline
Proto Copying & Optimizations to reduce copying expensive protocol buffers. \\
\hline
Protos & Inefficient use of protos. \\
\hline
Redundant Map Operations & See Figure~\ref{fig:dataset:anti-patterns:b}.  \\
\hline
Regexes & Inefficient use of regexes. \\
\hline
Return Value Optimization & Eliminate unnecessary copies of objects when returning from a function. \\
\hline
Salted Hash Sharding & To mitigate thread contention associated with a large shared table, shard the table and utilize a salted hash function to determine the appropriate shard for each key. \\
\hline
Specialized Swaps & Define a swap function in the class to swap objects of the class efficiently, and swap the object via std::swap within a scope guard. \\
\hline
String Comparison & String finds that can be replaced with more efficient finding functions. \\
\hline
String Concatenation & String concatenation that can be replaced with more efficient concatenation functions. \\
\hline
String Finds with Char & Use the character overload of  \verb|std::string::find()| and friends when the needle is a single character string literal. \\
\hline
String Views & String references that can be stored more efficiently. \\
\hline
Suboptimal Compressions & Changing or updating compression algorithms to improve performance. \\
\hline
Threads & Inefficient use of threads. \\
\hline
Type Erased Views & Use reference-like classes that do not own or copy data to enable code interaction with maps, random access ranges, and function objects without the necessity of templates. \\
\hline
Unmoved Unique Pointers & Avoid using an \verb|std::unique_ptr| if it is never std::move to or from another pointer. \\
\hline
Unnecessary Allocations & Allocation of a new object when reusing an existing object is possible. \\
\hline
Unnecessary Anys & Use of an inefficient any which requires runtime type information and should be avoided in favor of std::variant. \\
\hline
Unnecessary Bound Checking & \verb|std::vector::at| operator adds bounds checking overhead which can be avoided when the index is known to be valid, for example, part of a loop from 0 to size(). \\
\hline
Unnecessary Copied Args & Using a particular internal API causing unnecessary copies of arguments from a widely used internal framework. \\
\hline
Unnecessary Copies & Replacing variables with references. \\
\hline
Unnecessary Indirections & Reduce pointer chasing and other forms of indirection \\
\hline
Unnecessary Map Copies & Use of an inefficient mapping function for data processing causing unnecessary copies to occur. \\
\hline
Unnecessary Sorts & Use of a sorted or stable map/set when not required. \\
\hline
Vector Reads & Missed use of a read-only view to the elements of \verb|std::vector<T>|. \\
\bottomrule
\end{tabular}
\end{table}

\pagebreak
\clearpage
\section{LLM Trajectory Examples}
\label{sec:supplement:llm-trajectories}

This section provides examples of end-to-end LLM trajectories using the various prompting mechanisms described in Section~\ref{sec:edit-gen}. The prompts in blue show the query that is given to the LLM, and the LLM's responses are shown in grey, with additional syntax highlighting (added by us) to make the generated diff more readable. In the prompts, we added a placeholder for the target code that the prompt is applied to (the responses are based on calling the prompt with a representative example input).

\subsection{Zero-Shot}

This prompt format only includes the requested optimization, as well as the input code that it should be applied to. The model will respond with a snippet, based on its training.

\lstPrompt
\begin{lstlisting}[
    label={lst:zero_shot_prompt},
    caption={Example zero-shot prompt.},
    language=prompt
]
Reuse object allocations when possible to avoid allocating in a loop
{ target snippet }
\end{lstlisting}
\lstCodeDiff
\begin{lstlisting}[
    label={lst:zero_shot_diff},
    caption={Zero-shot generated code change.},
    language=code
]
@@■■■■■■■■■■■■■■■@@

       }
     }
 
+    ■■■■■■■■■■■■ key;
     for (const auto& [■■■■■■■■■■■■■■, value] : ■■■■■■■■■■■■■_map) {
-      ■■■■■■■■■■■■ key;
       key.ParseFromString(■■■■■■■■■■■■■■);
       emit_fn.Emit(key, value);
     }
\end{lstlisting}

\subsection{Few-Shot}

Here, we provide the model with specific positive and negative examples of an optimization. Just like before, the prompt then closes with the input code to optimize, and the model outputs a code diff when prompted.

\lstCodeDiff
\begin{lstlisting}[
    label={lst:few_shot_prompt},
    caption={Example few-shot prompt.},
    language=prompt
]
**Task**: As a C++ expert, you are required to analyze the use of the reserve method for std::vector. Ensure that vectors are pre-allocated ONLY when the size can be statically computed in advance.
**Guidelines**:
1. Avoid using reserve for trivial sizes; i.e., don't reserve space for just a few elements.
2. Avoid possible overallocations.
3. Consider special scenarios where reserve might be unnecessary or problematic.
4. The goal is to optimize memory usage without sacrificing runtime performance.
**Negative Examples**:
- **Example**:
   ```cpp
   std::vector<AutomatedRuns> failed_runs;
   std::vector<AutomatedRuns> successful_runs;
   successful_runs.reserve(Runs.size());
   failed_runs.reserve(Runs.size());
   for (auto Run : Runs) {
     if (Run.status == ■■■■■■■■■■■■■■■■■■■■■■■■■■■■■■■■■■■■■■■::SUCCESS) {
       successful_runs.push_back(Run);
     } else {
       failed_runs.push_back(Run);
     }
   }
   ```
   **Issue**: Overallocation by reserving space for all Runs.
- **Example**:
   ```cpp
   ■■■■■■■■■■■■■■■■■■■■■■■■■■■■■■.reserve(■■■■■■■■■■■■■■■■■■■■■■.size());
   for (const (■■■■■■■■■■■■■■■■■■■■■■■■■& ■■■■■■■■■■■■■■■■■■■■ :
   ■■■■■■■■■■■■■■■■■■■■■■) {
     if (■■■■■■■■■■■■■■■■■■■■.date >= start_date &&
     ■■■■■■■■■■■■■■■■■■■■.date <= end_date) {
       ■■■■■■■■■■■■■■■■■■■■■■■■■■■■■■.push_back(■■■■■■■■■■■■■■■■■■■■);
     }
   }
   return ■■■■■■■■■■■■■■■■■■■■■■■■■■■■■■■
   ```
   **Issue**: Overallocation by reserving space for all
   ■■■■■■■■■■■■■■■■■■■■■■.
- **Example**:
   ```cpp
   auto arg_landscape = ■■■■■■■■■■■■■■■■■■■■■■■■■->Compute(context);
   arg_landscape.Reserve(state_.responses.size());
   ```
   **Issue**: arg_landscape was already assigned at this point.
- **Example**:
   ```cpp
   ■■■■■■■■■■■■■■■■■■■■■■■■■■■■■■■■■■ ■■■■■■■■■■■■■■■■■■■■;
   ■■■■■■■■■■■■■■■■■■■■.reserve(options.■■■■■■■■■■■■■■■■■■■■■■■■■■■■■■);
   ■■■■■■■■■■■■■■■■■■■■■■■■■■■■■■■■■■■■■■■■■■■■■■■■■■■■
   ■■■■■■■■■■■■■■■■■■false,
   ■■■■■■■■■■■■■■■■■■■■■■■■■■■■■■■■■■■■■■■■■■■■■■■■■■■■■■■■■■■,
   ■■■■■■■■■■■■■■■■■■■■■);
   if (■■■■■■■■■■■■■■■■■■■■.empty()) return;
   ```
   **Issue**: ■■■■■■■■■■■■■■■■■■■■■■■■■■■■ clears the output vector and calls reserve on it.
**Positive Examples**:
- **Example**:
   ```cpp
   std::vector<std::pair<int, int>> ranges;
   ranges.reserve(num_ranges);
   //... (rest of the code)
   return ranges;
   ```
- **Example**:
   ```cpp
   std::vector<std::pair<int, int>> ranges;
   ranges.reserve(num_ranges);
   int last_added = 0;
   for (int i = 0; i < num_ranges; i++) {
     const int start = last_added;
     const int end = start + elements_per_range + (i < remainder ? 1 : 0);
     ranges.push_back(std::make_pair(start, end));
     last_added = end;
   }
   return ranges;
   ```
**Task**: Given the above examples and guidelines, determine if we can improve the following code by adding reserve:
{ target snippet }

\end{lstlisting}
\lstCodeDiff
\begin{lstlisting}[
    label={lst:few_shot_diff},
    caption={Few-shot generated code change.},
    language=code
]
@@■■■■■■■■■■■■■■■■@@
   absl::Status s;
   if (expr.union__size() > 0) {
     std::vector<■■■■■■■■■■■■■■■■■■■*> children;
+    children.reserve(expr.union__size());
     for (const auto& child : expr.union_()) {
       children.push_back(■■■■■■■■■■■■■■■■■■■■(child, parent_ref, context,
                                              relation_config, policy, source));
\end{lstlisting}

\clearpage

\subsection{Chain-of-Thought}

The key part of of this prompt if the last sentence – ``Let's think this through step by step''. This induces the model to generate a chain-of-thought sequence as opposed to directly outputting the answer.

\lstPrompt
\begin{lstlisting}[
    label={lst:cot_prompt},
    caption={Example chain-of-thought prompt.},
    language=prompt
]
You're a software engineer ■■■■■■■■■ and you have identified the destructor associated with { proto name } in { class name } to be very expensive.
Optimize { class name } by using proto arenas. There are following rules to follow:
1. Declare an instance of proto2::Arena before any instance of { proto name }.
2. Replace any instance of { proto name } with proto2::ArenaSafeUniquePtr<{ proto name }>
3. Every object associated with proto2::ArenaSafeUniquePtr<{ proto_name }> would be constructed using proto2::MakeArenaSafeUnique<{ proto name }>.
4. Ensure that the same arena object is shared across the different instances of proto2::ArenaSafeUniquePtr<{ proto name }>.
5. Use pointer semantics for every object associated with proto2::ArenaSafeUniquePtr<{ proto name }>.

{ target snippet }

Let's think this through step by step.
\end{lstlisting}

\lstCodeDiff
\begin{lstlisting}[
    label={lst:cot_diff},
    caption={Chain-of-thought generated code change.},
    language=code
]
Okay, let's analyze the provided code and apply the requested
changes to optimize ■■■■■■■■■■■■■■■■■■ using proto arenas.

**Understanding the Problem**

The destructor for ■■■■■■■■■■■■■■■■ is expensive, indicating
that it is not being allocated in an efficient way. Proto arenas
are a way to manage the allocation and deallocation of protocol
buffer messages in a more efficient way, especially in
high-throughput applications.

**Code Changes**

Here's the diff of the necessary code changes in
■■■■■■■■■■■■■■■■■■■■■■■■■■■■■■■■■:

**Explanation of Changes:**

1.  **Include arena_safe_unique_ptr.h:** We need to include the
header file that provides proto2::ArenaSafeUniquePtr.
2.  **Declare proto2::Arena:** We declare an proto2::Arena
object named arena_ before the declaration of ■■■■■■■■■■■■■■■■.
This arena object will be used to allocate the ■■■■■■■■■■■■■■
proto.
3.  **Replace ■■■■■■■■■■■■■■■■■■■■■■■■■■■■■■■■ with
proto2::ArenaSafeUniquePtr<■■■■■■■■■■■■■■■■■■■■■■■■■■■■■■■■>:**
The ■■■■■■■■■■■■■■■■ member is now of type
proto2::ArenaSafeUniquePtr<■■■■■■■■■■■■■■■■■■■■■■■■■■■■■■■■>.
4.  **Allocate ■■■■■■■■■■■■■■■■ on the arena:** We use
proto2::MakeArenaSafeUnique<■■■■■■■■■■■■■■■■■■■■■■■■■■■■■■■■>(
&arena_) to allocate the ■■■■■■■■■■■■■■ proto on the arena when
setting or swapping a new ■■■■■■■■■■■■■■.
5.  **Use pointer semantics:** We use pointer semantics (* or ->) 
to access the underlying ■■■■■■■■■■■■■■ proto. We now check
if the pointer ■■■■■■■■■■■■■■■■ is not null and use the pointer
to access the proto.
6.  **Clear ■■■■■■■■■■■■■■■■:** Use .reset() instead of =
std::nullopt to clear the ■■■■■■■■■■■■■■■■ pointer.

**Impact of Changes**

*   **Improved destructor performance:** The destructor of
■■■■■■■■■■■■■■■■■■ will be faster because the arena will
deallocate all the allocated ■■■■■■■■■■■■■■ objects at once,
avoiding the need to recursively call destructors.
*   **Potentially reduced memory fragmentation:** Arena
allocation can reduce memory fragmentation, as all
arena-allocated objects are released together.
*   **Reduced allocation overhead:** Arena-based allocation is
faster than standard new and delete.

**Additional Notes**

*   It is important to note that if ■■■■■■■■■■■■■■■■■■ is used
in a context where the ■■■■■■■■■■■■■■ is frequently set and
cleared, then the arena might grow significantly, and it might
be advantageous to reset the arena periodically.
*   The provided solution assumes that ■■■■■■■■■■■■■■■■■■ does
not need to return a raw pointer to the ■■■■■■■■■■■■■■. If this
is needed, then it is necessary to return the raw pointer from
the ■■■■■■■■■■■■■■■■ member.

This approach leverages the efficiency of proto arenas to
optimize the performance of ■■■■■■■■■■■■■■■■■■ and its handling
of the ■■■■■■■■■■■■■■■■ member.


@@■■■■■■■■■■■■■■■■@@
 #include "■■■■■■■■■■■■■■■■■■/public/callbackmetric.h"
 #include "■■■■■■■■■■■■■■■■■■/public/concurrent.h"
 #include "■■■■■■■■■■■■■■■■■■/public/counter.h"
+#include "■■■■■■■■■■/arena/arena_safe_unique_ptr.h"
 #include "■■■■■■■■■■/proto/descriptor.proto.h"
 #include "■■■■■■■■■■/public/descriptor.h"
 #include "■■■■■■■■■■/public/dynamic_message.h"
@@■■■■■■■■■■■■■■■■@@
   }
 
   bool ■■■■■■■■■■■■() const override {
-    return ■■■■■■■■■■■.has_value();
+    return ■■■■■■■■■■■ != nullptr;
   }
 
   const ■■■■■■■■■■■■■■■■■■■■■■& ■■■■■■■■■() const override {
-    if (■■■■■■■■■■■.has_value()) {
+    if (■■■■■■■■■■■ != nullptr) {
       return *■■■■■■■■■■■;
     }
     ProgrammingErrorDfatal()
@@■■■■■■■■■■■■■■■■@@
 
   void Set■■■■■■■■■(
       const ■■■■■■■■■■■■■■■■■■■■■■& ■■■■■■■■■■) override {
-    ■■■■■■■■■■■ = ■■■■■■■■■■;
+    if (■■■■■■■■■■■ == nullptr) {
+      ■■■■■■■■■■■ =
+          proto2::MakeArenaSafeUnique<■■■■■■■■■■■■■■■■■■■■■■>(
+              &arena_);
+    }
+    *■■■■■■■■■■■ = ■■■■■■■■■■;
   }
 
   void Swap■■■■■■■■■(■■■■■■■■■■■■■■■■■■■* ■■■■■■■■■■) override {
     if (■■■■■■■■■■ == nullptr) {
       ProgrammingErrorDfatal()
-          << __LOG__ << " passed `■■■■■■■■■■` is a nullptr";
+          << __LOG__ << " passed `■■■■■■■■■■` is a nullptr.";
       return;
     }
-    if (!■■■■■■■■■■■.has_value()) {
-      ■■■■■■■■■■■.emplace();
-    }
-    ■■■■■■■■■■■->Swap(■■■■■■■■■■);
+    if (■■■■■■■■■■■ == nullptr) {
+      ■■■■■■■■■■■ =
+          proto2::MakeArenaSafeUnique<■■■■■■■■■■■■■■■■■■■■■■>(
+              &arena_);
+    }
+    ■■■■■■■■■■■->Swap(■■■■■■■■■■);
   }
 
   void Clear■■■■■■■■■■■■■■■■■■■■■■() override {
-    ■■■■■■■■■■■ = std::nullopt;
+    ■■■■■■■■■■■.reset();
   }
 
   int32_t ■■■■■■■■■■■■■_count() const override { return 1; }
@@■■■■■■■■■■■■■■■■@@
   ■■■■■■■■■■■■■■■■■■■■■■■■■■■■■■■■■■■■■■■■■■■■■■■■■■■■■■■■
   stats_census::TagSetProtoWithAdditionalMetadata
       census_tagset_proto_with_metadata_;
-
+  
+  proto2::Arena arena_;
-  std::optional<■■■■■■■■■■■■■■■■■■■■■■> ■■■■■■■■■■■;
+  proto2::ArenaSafeUniquePtr<■■■■■■■■■■■■■■■■■■■■■■> ■■■■■■■■■■■;
 };
\end{lstlisting}

\clearpage

\subsection{ReAct}

When using ReAct, the LLM is invoked multiple times. In each step, it is given the full trajectory of the execution so far. The given prompt format will induce the model to follow a particular structure that results in it generating a ``Thought'' followed by an ``Action'', which is an executable command. The model knows this structure from its pre-training data, which included ReAct examples (an alternative approach would be to provide examples of the desired structure as few-shot examples in the prompt itself). The action is then executed, the output of the action is copied back into the prompt, and the next iteration begins. The final action results in the generation of a diff.

\lstPrompt
\begin{lstlisting}[
    label={lst:react_prompt},
    caption={Example ReAct prompt.},
    language=prompt
]
# Instruction:
Make a performance enhancing code edit that accomplishes the following:
```txt
Restricts the lambda capture scope
```
# Thought:
We need to first take a look at the source code.
# Action:
cat { target path }
# Observe:
```C++
{ target snippet }
```
# Thought:
Now we can write a code diff that will improve the source code's performance.
# Action:
\end{lstlisting}
\lstCodeDiff
\begin{lstlisting}[
    label={lst:react_diff},
    caption={ReAct generated code change.},
    language=code
]
@@■■■■■■■■■■■■■■■■@@

   for (auto& query : *mutable_queries) {
     const int ■■■■■■■■■■■■■■■■■■■ = query.■■■■■■■■■();
     proto2::util::RemoveIf(
-        query.■■■■■■■■■■■■(), [this](const ■■■■■■■■■■■■■■■■* ■■■) {
+        query.■■■■■■■■■■■■(),
+        [slot_position_threshold =
+             slot_position_threshold_](const ■■■■■■■■■■■■■■■■* ■■■) {
           return (■■■->is_pla() &&
-                  ■■■->slot_position() > slot_position_threshold_);
+                  ■■■->slot_position() > slot_position_threshold);
         });
     const int ■■■■■■■■■■■■■■■■■ = query.■■■■■■■■■();
     ■■■■■■■■■■■■■■■■■■■■■■■■■■■■■■■■->IncrementBy(■■■■■■■■■■■■■■■■■■■ -
\end{lstlisting}
\clearpage
\section{Multi-File Edits}
\label{sec:supplement:multi-file-edits}

Although the bulk of this paper uses examples with single-function edits, \name{} is not restricted to function-level edits. \name{} uses an LLM fine-tuned with multi-file code changes and the techniques used in \name{} apply to multi-file edits as well: e.g., localizing through code embeddings; targeted prompting for generating code optimizations; and validation through testing, self-review, and code reviews. 

We recently expanded \name{} to apply more multi-file code changes and we have found that the primary difference between multi-file and function-level edits is a necessity for descriptive prompting. For example, one of our anti-patterns involves optimizing usage of protocol buffers~\cite{protobuf}. These protocol buffer edits are particularly complex because they can involve changes across multiple files (often >5 files at a time). Providing more detailed instructions in the prompts for these edits allows us to ensure that all multi-file dependencies/references are captured in the generated edit.

Prompt~\ref{lst:cot_prompt} shows a prompt that allowed us to effectively generate multi-file protocol buffer edits, and Code Diffs~\ref{lst:protobuf_change_h} and~\ref{lst:protobuf_change_cc} show edits generated with the prompt.

\lstCodeDiff
\begin{lstlisting}[
    label={lst:protobuf_change_h},
    caption={File 1 of a change generated with Prompt~\ref{lst:cot_prompt}.},
    language=code
]
@@■■■■■■■■■■■■■@@

 #include "■■■■■■■■■■■■■■■■■■■■■■■■■■■■■■■"
 #include "■■■■■■■■■■■■■■■■■■■■■■■■■■■■■■■■■■■■■"
 #include "■■■■■■■■■■■■■■■■■■■■■■■■■■■■"
+#include "■■■■■■■■■■■■■■■■■■■■■■■■■■■■"
 #include "■■■■■■■■■■■■■■■■■■■■■■■■■■■■■■■■■"
 
 namespace ■■■■■■■ {
@@■■■■■■■■■■■■■■■■■@@

 
   ■■■■■■■■■■■■■■■■■■■■■■■■■■■■■■■■■■■■■■■■■■■■■■■■■■■■■■
   const MutationProto* mutation() const {
-    return (mutation_ != nullptr) ? mutation_.get() : &entry_.mutation();
-  }
-
-  bool is_■■■■_entry() const { return entry_.header().has_■■■■_id(); }
-  DatabaseId database_id() const { return entry_.header().database_id(); }
-  int32_t ■■■■_id() const { return entry_.header().■■■■_id(); }
+    return (mutation_ != nullptr) ? mutation_.get() : &entry_->mutation();
+  }
+
+  bool is_■■■■_entry() const { return entry_->header().has_■■■■_id(); }
+  DatabaseId database_id() const { return entry_->header().database_id(); }
+  int32_t ■■■■_id() const { return entry_->header().■■■■_id(); }
   LogEntryHeaderProto* mutable_entry_header() {
-    return entry_.mutable_header();
+    return entry_->mutable_header();
   }
   bool is_■■■■_info_persisted_in_entry_header() const {
-    return entry_.header().has_checksummed_■■■■_encryption_info();
+    return entry_->header().has_checksummed_■■■■_encryption_info();
   }
 
   ■■■■■■■■■■■■■■■■■■■■■■■■■■■■■■■■■■■■■■■■■■■■■■■■■■■■■■■■■■■■■■■■■■
@@■■■■■■■■■■■■■■■@@

   const std::string database_monitoring_name_;
 
   perftools::tracing::OperationTraceSpan trace_span_;
-  LogEntryProto entry_;
+
+  proto2::Arena arena_for_entry_;
+  proto2::ArenaSafeUniquePtr<LogEntryProto> entry_;
 
   std::string serialized_header_;
\end{lstlisting}
\begin{lstlisting}[
    label={lst:protobuf_change_cc},
    caption={File 2 of a change generated with Prompt~\ref{lst:cot_prompt}.},
    language=code
]
@@■■■■■■■■■■■■■@@

 #include "■■■■■■■■■■■■■■■■■■■■■■■■■■■■■■■■■■■■■■■■■"
 #include "■■■■■■■■■■■■■■■■■■■■■■■■■■■■■■■■■■■■■■■■■■■■■■■■■■"
 #include "■■■■■■■■■■■■■■■■■■■■■■■■■■■■■■■■"
+#include "■■■■■■■■■■■■■■■■■■■■■■■■■■■■■■■■■■■■■■■■"
+#include "■■■■■■■■■■■■■■■■■■■■■■■■■■■■■■■■■■■■■■■■■■■■■"
 #include "■■■■■■■■■■■■■■■■■■■■■■■■"
 #include "■■■■■■■■■■■■■■■■■■■■■■"
 #include "■■■■■■■■■■■■■■■■■■■■■■■■■■■■■■■■■■■■■■■"
@@■■■■■■■■■■■■■@@

 #include "■■■■■■■■■■■■■■■■■■■■■■■■■■■■■■■■"
 #include "■■■■■■■■■■■■■■■■■■■■■■■■■"
 #include "■■■■■■■■■■■■■■■■■■■■■■■■■■■■■■■"
+#include "■■■■■■■■■■■■■■■■■■■■■■■■■■■■■■■■■■■■■■■■■■■■■■■■■"
 #include "■■■■■■■■■■■■■■■■■■■■■■■■■"
 #include "■■■■■■■■■■■■■■■■■■■■■■■"
 #include "■■■■■■■■■■■■■■■■■■■■■■■■■■■■■■■■■■■■■■"
@@■■■■■■■■■■■■■■■@@

       prev_sequence_(prev_sequence),
       tablet_uid_(tablet_uid.Encode()),
       durable_cb_(durable_cb) {
-  ■■■■■■■■■■■■■■■■■■■* header = entry_.mutable_header();
+  entry_ = proto2::MakeArenaSafeUnique<■■■■■■■■■■■■■>(&arena_for_entry_);
+  ■■■■■■■■■■■■■■■■■■■* header = entry_->mutable_header();
   header->set_prev_sequence(prev_sequence);
   header->set_database_id(database_id);
   header->set_tablet_uid(tablet_uid_);
@@■■■■■■■■■■■■■■■@@

   }
 
   FillTabletVersion(header->mutable_version());
-  entry_.set_allocated_mutation(mutation.release());
-  entry_.set_creation_wall_time_micros(■■■■■■■■■■■■■■■■■■(creation_time_));
+  proto2::SetMessageField<"mutation">(std::move(mutation), *entry_);
+  entry_->set_creation_wall_time_micros(■■■■■■■■■■■■■■■■■■(creation_time_));
   if (isolation_info) {
     isolation_info_ = *isolation_info;
     isolation_info_->SetDatabaseMonitoringName(database_monitoring_name_);
@@■■■■■■■■■■■■■■■■■@@

 }
 
 void ■■■■■■■■■■■■■■■■■■■::Finalize(int64_t ■■■■■■■■■■■■■■■) {
-  entry_.mutable_header()->set_sequence(■■■■■■■■■■■■■■■);
+  entry_->mutable_header()->set_sequence(■■■■■■■■■■■■■■■);
   sequence_ = ■■■■■■■■■■■■■■■;
 }
 
 void ■■■■■■■■■■■■■■■■■■■::SetWriteTime(■■■■■■■■■■ time) {
   ■■■■■■■■■■■■■■■■■■■■■■■■■■■■■■■■■■■■■■■■■■■■■■■
-  entry_.set_wall_time_micros(■■■■■■■■■■■■■■■■■■(time));
+  entry_->set_wall_time_micros(■■■■■■■■■■■■■■■■■■(time));
 }
 
 void ■■■■■■■■■■■■■■■■■■■::SetFlushTime(■■■■■■■■■■ time) { flush_time_ = time; }
@@■■■■■■■■■■■■■■■■■@@

 const ■■■■■■■■■■■■■& ■■■■■■■■■■■■■■■■■■■::contents() const {
   if (should_encrypt_entry_) {
     ■■■■■■■■■■■■■■■■■■■■■■■■■■■■■■■■■■■■■■■■
-    CHECK(!entry_.■■■■■■■■■■■■■■);
-    CHECK(entry_.■■■■■■■■■■■■■■■■■■■■■■■■);
-  }
-  return entry_;
+    CHECK(!entry_->■■■■■■■■■■■■■■);
+    CHECK(entry_->■■■■■■■■■■■■■■■■■■■■■■■■);
+  }
+  return *entry_;
 }
 
 ■■■■■■■■■■ ■■■■■■■■■■■■■■■■■■■::AppendLogEntryToBuffer(LogRecordBuffer& buffer) {
   if (!should_encrypt_entry_) {
     ■■■■■■■■■■■■■■■■■■■■■■■■■■■■■■■■■■■■■■■■■■■■■■■■■■■■■■■■■■■■■■■■■
     ■■■■■■■■■■■■■■■■■
-    return buffer.Append(entry_, ■■■■■■■■■■■■■::kChecksumFieldNumber);
+    return buffer.Append(*entry_, ■■■■■■■■■■■■■::kChecksumFieldNumber);
   } else {
     ■■■■■■■■■■■■■■■■■■■■■■■■■■■■■■■■■■■■■■■■■■■■■■■■■■■■■■■■■■■■■■■■■■■■
     ■■■■■■■■■■■■■■■■■■■■■■■■■■■■■■■■■■■■■■■■■■■■■■■■■■■■■■■■■■■■■■■
     CHECK(!serialized_header_.empty());
     ScopedEncryptedLogEntryForSerialization to_be_serialized(
-        &entry_, &serialized_header_);
+        entry_.get(), &serialized_header_);
     return buffer.Append(to_be_serialized.MessageToSerialize(),
                          ■■■■■■■■■■■■■::kChecksumFieldNumber);
   }
@@■■■■■■■■■■■■■■■■@@

                             const ■■■■■■■■■■& mutation_cord) {
   if (should_encrypt_entry_) {
     ■■■■■■■■■■■■■■■■■■■■■■■■■■■■■■■■■■■■■■■■■■■■■■■■■■■■■■■■■■■■■■■■■■■■■■■■■■
-    entry_.header().SerializeToString(&serialized_header_);
+    entry_->header().SerializeToString(&serialized_header_);
     ■■■■■■■■■■■■■■■■■■■■■■■■■■■■■■■■■■■■■■■■
-    mutation_ = ■■■■■■■■■■■■■■■■(&entry_, mutation_cord, serialized_header_);
+    mutation_ =
+        ■■■■■■■■■■■■■■■■(entry_.get(), mutation_cord, serialized_header_);
   }
 }
\end{lstlisting}

\clearpage
\section{Self-Validation}
\label{sec:supplement:self-validation}

We provide an example of LLM self-validation for edits generated to fix missing vector reserves (Figure~\ref{fig:dataset:anti-patterns:c}). For evaluation of these edits, we prompt the LLM to answer several questions:

\begin{enumerate}
    \item Is the reserved amount in the vector appropriate?
    \item Is the \verb|push_back| of elements in the vector  happening within the scope of a conditional block like \verb|if|? (We want to avoid reserving when we cannot pre-determine the vector size.)
    \item Is the vector being modified in the surrounding code and are there \verb|push_back|s happening outside the context of a for-loop? (If yes, we want to avoid reserving to avoid miscalculations.)
    \item Does the reserve count in vector depend on nested for-loops? (We want to avoid reserving when we cannot easily compute the nested loop values.)
\end{enumerate}

\noindent Prompt~\ref{lst:self_eval_question} provides an example of a self-validation prompt template for the first question. When given Code Snippet~\ref{lst:self_eval_function_1}, the LLM validation returns ``No'' as shown in Response~\ref{lst:self_eval_response_1}. An example code snippet that receives the opposite response (``Yes'') is shown in Code Snippet~\ref{lst:self_eval_function_2} and Response~\ref{lst:self_eval_response_2}.

\lstPrompt
\begin{lstlisting}[
    label={lst:self_eval_question},
    caption={Example of a self-validation prompt.},
    language=prompt
]
You are an expert C++ performance engineer tasked with reviewing code. Your goal is to assess the quality of a code snippet which seeks to preallocate a vector to avoid reallocations when elements are continuously added in a for loop. You will be presented with a code and a question.

  Code:
  ```cpp
  { target snippet }
  ```

  Question:
  Is the reserved amount in the vector { target vector } appropriate (only reserve when we know the exact size in advance)? (Yes or No)
\end{lstlisting}
\lstCode
\begin{lstlisting}[
    label={lst:self_eval_function_1},
    caption={Example input code for Prompt~\ref{lst:self_eval_question}.},
    language=code
]
■■■■■■■■■■■■■■<std::vector<std::string>> ■■■■■■■■■■■■■■■■■■■■■■■■(
    const ■■■■■■■■& ■■■■■■■■■, ■■■■■■■■■■<const std::string> input_full_names) {
  ■■■■■■■■■■■■■■■■■■■
  std::vector<std::string> input_paths;
  input_paths.reserve(input_full_names.size());
  std::vector<std::string> missing_full_names;
  ■■■■■■■■■■■■■■■■■■■■■■■■■■■■■■■■■■■■■■■■■■■■■■■■■■■■■■■■■■■■■
  ■■■■■■■■■■■■■■■■■■■<std::string, std::string> ■■■■■■■■■■■■■■■■■■■■■■_map;
  for (const auto& data : ■■■■■■■■■.inputs()) {
    ■■■■■■■■■■■■■■■■■■■■■■_map[data.full_name()] = data.path();
  }
  ■■■■■■■■■■■■■■■■■■■■■■■■■■■■■■■■■■■■■■■■■■■■■■■■■■■■■■■■■■■■■
  for (const std::string& input_full_name : input_full_names) {
    if (auto it = ■■■■■■■■■■■■■■■■■■■■■■_map.find(input_full_name);
        it != ■■■■■■■■■■■■■■■■■■■■■■_map.end()) {
      input_paths.push_back(it->second);
    } else {
      missing_full_names.push_back(input_full_name);
    }
  }
  ■■■■■■■■■■■■■■■■■■■■■■■■■■■■■■■■■■■■■■■■■■■■■
  if (!missing_full_names.empty())
    return ■■■■■■■■■■■■■■Error(
        ■■■■■■■■■■■■("No paths found in input data for below full names: ",
                     ■■■■■■StrJoin(missing_full_names, ",")));
  ■■■■■■■■■■■■■■■■■■■■■■■
  return input_paths;
}
\end{lstlisting}

\clearpage
\lstResponse
\begin{lstlisting}[
    label={lst:self_eval_response_1},
    caption={LLM response to Prompt~\ref{lst:self_eval_question} with input Code~\ref{lst:self_eval_function_1}.},
    language=llm
]
Okay, let's analyze the `reserve` call in the provided C++ code snippet.

**Analysis:**

The code aims to find matching paths from a `■■■■■■■■` based on a list of `input_full_names`. It builds a hash map for quick lookup and then iterates through the `input_full_names`. If a name is found in the map, its corresponding path is added to the `input_paths` vector. If not, the name is added to a `missing_full_names` vector, and if this vector is non-empty at the end of the iteration, an error will be returned.

The relevant line is:

```c++
input_paths.reserve(input_full_names.size());
```

Here's a breakdown of whether the `reserve` is appropriate:

*   **Worst-Case Scenario:** The code reserves space for `input_full_names.size()` elements. This is the *maximum* number of paths that could be added to `input_paths`.  This happens only when all `input_full_names` exist in `■■■■■■■■■.inputs()`.
*   **Likely Scenario:**  It's possible (and perhaps even probable) that some of the `input_full_names` will *not* be found in the `■■■■■■■■■■■■■■■■■■■■■■_map`.  In this case, the `reserve` call will have allocated *more* space than necessary.
*    **Exact size unknown**: We do not know the exact size of the vector in advance.

**Conclusion and Explanation:**
The `reserve` call is *not perfectly appropriate* because it reserves the maximum possible size, not the *exact* size. The exact size is not known beforehand, because not all names in `input_full_names` are guaranteed to be found in `■■■■■■■■■.inputs()`.  A more precise `reserve` would be impossible without iterating through `input_full_names` and checking the map first, which would duplicate the work of the main loop.

The current `reserve` is a reasonable optimization, as it avoids reallocations in the best-case/worst-case scenario where all names are found. However, it may lead to over-allocation in other cases.  The performance benefit of the current approach likely outweighs the potential cost of over-allocation in most realistic scenarios, especially if the expected number of missing names is small compared to the total number of names.  If we knew, for instance, that 99% of names were always present, the current approach would be very efficient.

**Answer:**

No
\end{lstlisting}
\lstCode
\begin{lstlisting}[
    label={lst:self_eval_function_2},
    caption={Example input code for Prompt~\ref{lst:self_eval_question}.},
    language=code
]
std::vector<■■■■■■■■■■■■■■■■■■■■■■■■■■■■>::■■■■■■■■■■■■■■■■■■>
■■■■■■■■■■■■■■■■■■■■■■■(const std::vector<int> criterions) {
  std::vector<■■■■■■■■■■■■■■■■■■■■■■■■■■■■::■■■■■■■■■■■■■■■■■■> criterion;
  criterion.reserve(criterions.size());
  for (auto id : criterions) {
    ■■■■■■■■■■■■■■■■■■■■■■■■■■■■::■■■■■■■■■■■■■■■■■■ tc;
    tc.mutable_criterion()->set_criterion_id(id);
    criterion.push_back(tc);
  }
  return criterion;
}
\end{lstlisting}

\clearpage
\lstResponse
\begin{lstlisting}[
    label={lst:self_eval_response_2},
    caption={LLM response to Prompt~\ref{lst:self_eval_question} with input Code~\ref{lst:self_eval_function_2}.},
    language=llm
]
Okay, let's analyze the code snippet and answer the question.

**Code Analysis**

The code aims to create a vector of `■■■■■■■■■■■■■■■■■■■■■■■■■■■■::■■■■■■■■■■■■■■■■■■` objects.  It takes a vector of integers (`criterions`) as input.  For each integer in the input vector, it creates a `■■■■■■■■■■■■■■■■■■` object, sets its `criterion_id`, and then adds it to the output vector `criterion`. The core of the question is the use of `criterion.reserve(criterions.size());`.

**Explanation and Answer**

*   **`reserve()` vs. `resize()`:**  It's crucial to distinguish between `reserve()` and `resize()`.
    *   `reserve(n)`:  Allocates enough memory to hold `n` elements *without* changing the size of the vector (i.e., `vector.size()` remains unchanged).  Future `push_back()` operations (up to `n` elements) will not trigger reallocations, which are expensive.
    *   `resize(n)`: Changes the size of the vector to `n`.  If `n` is larger than the current size, the vector is expanded, and new elements are value-initialized (or default-constructed for objects). If `n` is smaller, the vector is truncated.

*   **Appropriateness of `reserve()` in this case:** The code correctly uses `reserve(criterions.size())`.  The loop iterates exactly `criterions.size()` times, and each iteration adds one element to the `criterion` vector. Therefore, we *know* the final size of the `criterion` vector in advance: it will be equal to `criterions.size()`.  Using `reserve()` is the optimal approach here.  It prevents any reallocations during the loop, leading to better performance, especially when `criterions.size()` is large.

*   **Why not `resize()`?** Using `resize(criterions.size())` would be *incorrect* and potentially less efficient. `resize()` would create `criterions.size()` default-constructed `■■■■■■■■■■■■■■■■■■` objects *before* the loop. The loop would then *copy* into these existing objects. The initial default construction is wasteful, and depending on the compiler and standard library implementation, might even be *less* efficient than the reserve-then-copy approach, as resize can also call the copy constructor to grow the container.  `resize` also affects the `size()` of the vector, which is not the intent. We want the size to grow naturally with the `push_back` calls.

*   **Worst case scenario with no `reserve()`**: Without the `reserve()` call, the `std::vector` would start with a small allocated capacity.  As elements are added via `push_back()`, the vector would need to reallocate its internal storage multiple times. Each reallocation involves:
    1.  Allocating a new, larger memory block.
    2.  Copying all existing elements from the old memory block to the new one.
    3.  Deallocating the old memory block.
    This reallocation process is O(N) in time complexity, and if it happens repeatedly, the overall performance degrades to O(N^2) in the worst case (though `std::vector` implementations typically use an exponential growth strategy to mitigate this).

**Conclusion**

The use of `reserve()` is appropriate and beneficial in this code. We know the exact size of the vector in advance, and `reserve()` is used to allocate the memory upfront, preventing costly reallocations.

**Answer to the Question:**

Yes
\end{lstlisting}

\clearpage
\section{Interactions of \name{} and Code Review}
\label{sec:supplement:experirence-reports}

For every change generated by \name{}, a human code owner is asked to review the change before the change is submitted to production. For 40\%, 5\%, and 41\% of copy, map, and vector changes, respectively, \name{}-generated changes were directly approved and submitted to production without the human reviewers leaving any feedback that had to be resolved. Table~\ref{table:casestudies} in the main portion of the paper shows a breakdown of when additional action was required. We now provide a few examples of these specific reviewer interactions below.

\begin{figure}[htbp]
    \centering
    \begin{subfigure}[t]{0.65\linewidth}
        \centering
        \fbox{\includegraphics[width=1\linewidth]{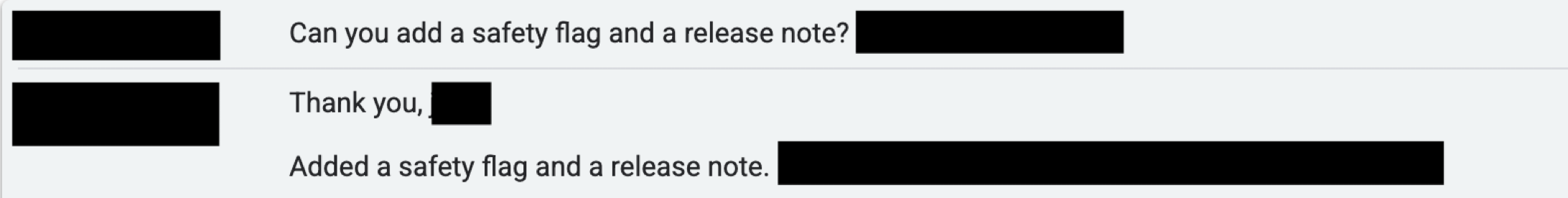}}
        \caption{Feedback that had to be resolved before \name{}'s change was submitted to production.}
        \label{fig:reviewer_feedback_1}
    \end{subfigure}
    \begin{subfigure}[t]{0.65\linewidth}
        \centering
        \fbox{\includegraphics[width=1\linewidth]{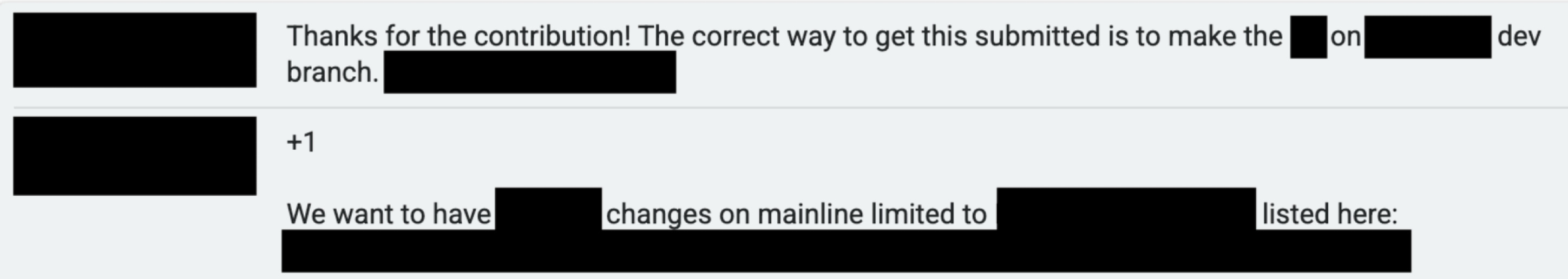}}
        \caption{Feedback rejecting change due to reasons unrelated to code performance.}
        \label{fig:reviewer_feedback_2}
    \end{subfigure}
    \caption{Examples of reviewer feedback.}
    \label{fig:reviewer_feedback}
\end{figure}

Oftentimes, users left comments that were extraneous to the efficiency optimization on the code change (\verb|S_USER| in Table~\ref{table:casestudies}). These comments had to be addressed before the code change was submitted to production. Figure~\ref{fig:reviewer_feedback_1} shows an example of feedback provided for a change that was submitted to production after reviewer comments were resolved. 

Other times, users rejected changes (\verb|R_USER| in Table~\ref{table:casestudies}) and the changes were not submitted to production. This could be due to the human reviewer judging the code change to actually not be performance enhancing or the human reviewer wanting the code to be untouched for other reasons (e.g., style preferences). Figure~\ref{fig:reviewer_feedback_2} provides an example of such feedback.

\name{} has also encountered code changes where feedback from human reviewers has caused regressions not initially present. For example, \name{} generated the change in Code Diff~\ref{lst:reverted_change:before}.
This change attempts to remove the redundant map lookups on \verb|opt_out_handler_|. The reviewer for this change suggested using \verb|try_emplace| instead, and in response to the reviewer’s comments, the code was modified as reflected in Code Diff~\ref{lst:reverted_change:after}.

However \verb|try_emplace| evaluates its argument expression every time, regardless of whether an insertion has occurred. This led to many calls to \verb|make_unique| and caused a large increase in CPU consumption. Due to the regression, the change had to be reverted (\verb|R_REVERT| in Table~\ref{table:casestudies}).

\clearpage

\lstCodeDiff
\begin{lstlisting}[
    label={lst:reverted_change:before},
    caption={Initial \name{}-generated code diff.},
    language=code
]
@@■■■■■■■■■■■■■■■■@@
     const ■■■■■■■■■■■■■■■■■& transform_task,
     const ■■■■■■■■■■■■■■■<int32_t>* ■■■■■■■■■■■■■■■■■■■) {
   transform_task_ = transform_task;
-  const ■■■■■■■■■■■■■■■<int32_t>* ■■■■■■■■■■ = (■■■■■■■■■■■■■■■■■■■ == nullptr)
-                                                   ? &■■■■■■■■■■■■■■■■■■■■■■■■
-                                                   : ■■■■■■■■■■■■■■■■■■■;
-  std::string handler_key = ■■■■■■■■■■■■■(*■■■■■■■■■■, "-");
-  if (opt_out_handler_.find(handler_key) == opt_out_handler_*.end()) {
-    opt_out_handler_[handler_key] =
-        std::make_unique<■■■■■■■■■■■■■■■■■■■■■■■■■■■■>(
-            *■■■■■■■■■■, opt_out_info_, transform_task_;
+  std::string handler_key =
+      ■■■■■■■■■■■■■((■■■■■■■■■■■■■■■■■■■ == nullptr) ? ■■■■■■■■■■■■■■■■■■■■■■■_
+                                                     : *■■■■■■■■■■■■■■■■■■■,
+                    "-");
+  auto& opt_out_handler = opt_out_handler_[handler_key];
+  if (opt_out_handler == nullptr) {
+    opt_out_handler = std::make_unique<■■■■■■■■■■■■■■■■■■■■■■■■■■■■>(
+        (■■■■■■■■■■■■■■■■■■■ == nullptr) ? ■■■■■■■■■■■■■■■■■■■■■■■■
+                                         : *■■■■■■■■■■■■■■■■■■■,
+        opt_out_info_, transform_task_);
   }
\end{lstlisting}
\lstCodeDiff
\begin{lstlisting}[
    label={lst:reverted_change:after},
    caption={\name{}-generated code diff after reviewer's feedback.},
    language=code
]
@@■■■■■■■■■■■■■■■■@@
     const ■■■■■■■■■■■■■■■■■& transform_task,
     const ■■■■■■■■■■■■■■■<int32_t>* ■■■■■■■■■■■■■■■■■■■) {
   transform_task_ = transform_task;
-  const ■■■■■■■■■■■■■■■<int32_t>* ■■■■■■■■■■ = (■■■■■■■■■■■■■■■■■■■ == nullptr)
-                                                   ? &■■■■■■■■■■■■■■■■■■■■■■■■
-                                                   : ■■■■■■■■■■■■■■■■■■■;
-  std::string handler_key = ■■■■■■■■■■■■■(*■■■■■■■■■■, "-");
-  if (opt_out_handler_.find(handler_key) == opt_out_handler_.end()) {
-    opt_out_handler_[handler_key] =
-        std::make_unique<■■■■■■■■■■■■■■■■■■■■■■■■■■■■>(
-            *■■■■■■■■■■, opt_out_info_, transform_task_);
-  }
+  std::string handler_key =
+      ■■■■■■■■■■■■■((■■■■■■■■■■■■■■■■■■■ == nullptr) ? ■■■■■■■■■■■■■■■■■■■■■■■_
+                                                     : *■■■■■■■■■■■■■■■■■■■,
+                    "-");
+  opt_out_handler_.try_emplace(
+      handler_key,
+      std::make_unique<■■■■■■■■■■■■■■■■■■■■■■■■■■■■>(
+          (■■■■■■■■■■■■■■■■■■■ == nullptr) ? ■■■■■■■■■■■■■■■■■■■■■■■■
+                                           : *■■■■■■■■■■■■■■■■■■■,
+          opt_out_info_, transform_task_));
\end{lstlisting}
\clearpage
\section{Reproducibility}
\label{sec:supplement:reproducibility}

Since \name{} is built for \google{}, the majority of \name{}-generated changes shown in this paper have been targeted at \google{}'s code base. However, \name{}'s techniques can be applied in the context of other organizations' systems infrastructure and code bases. Today, there are many open-source LLMs~\cite{deepseek-llm, grattafiori2024llama3herdmodels, jiang2023mistral7b} that can be fine-tuned with open-source libraries or with LLM offerings~\cite{nanoGPT, PEFT, torchtune, unsloth}. Our methods for identifying performance opportunities and localizing candidate code can also be replicated using various open-source performance profiling tools \cite{blanco2023practical} and similarity-search libraries like ScaNN~\cite{guo2020scann}, with required resources scaling with the size of the organization's code base.

Additionally, \name{} has generated code changes not only on \google{}'s code base but also on open-source code. Code Diff ~\ref{lst:open_source_diff} shows an example of a change generated by \name{} on TensorFlow (\url{https://github.com/tensorflow/tensorflow})~\cite{tensorflow2015}. This change removes a redundant map lookup and was committed into TensorFlow's upstream repository~\cite{tensorflowPR}.

\lstCodeDiff
\begin{lstlisting}[
    label={lst:open_source_diff},
    caption={Example of a code change generated on open-source code.},
    language=code
]
@@ -346,8 +346,9 @@

         fanouts_[output].emplace(node, -1);
       } else {
         max_input_port = i;
-        max_regular_output_port_[output.node] =
-            std::max(max_regular_output_port_[output.node], output.port_id);
+        int& max_regular_output_port = max_regular_output_port_[output.node];
+        max_regular_output_port =
+            std::max(max_regular_output_port, output.port_id);
         fanouts_[output].emplace(node, i);
       }
     }
\end{lstlisting}
\clearpage
\section{Comparison With Static Analysis}
\label{sec:supplement:comparision-with-clang-tidy}
\noindent Table~\ref{tab:comparison_clang_tidy_eco} describes 6 different common sub-optimal vector-reserve patterns and lists whether or not static analysis are able to resolve them. 5 out of the 6 are not resolvable with Clang-Tidy and none are optimized by CppCheck, whereas \name{} is able to optimize each one. 
The table also shows speedup results from \name{} optimizations on microbenchmarks, run against varying container sizes (from 8 to 4096 elements). 
Here we describe each vector-reserve pattern in more detail.

\niparagraph{Pattern A: Standard Loop Insertion}
Clang-Tidy and \name{} are both able to optimize loops using standard library vector insertions such as \texttt{.push\_back}.

\begin{figure}[H]
\centering
\lstCode
\begin{lstlisting}[
    label={lst:vector_reserve_pattern_a},
    caption={Optimization example for Pattern A.},
    language=code
]
dirs.reserve(groups_.size());
for (Group g : groups_) {
    dirs.push_back(GroupDir(g));
}
\end{lstlisting}
\end{figure}

\niparagraph{Pattern B: Bulk Move via Algorithms}
This pattern moves elements into a destination vector using an algorithm wrapper (e.g., \texttt{absl::c\_move} with \texttt{std::back\_inserter}).

\begin{figure}[H]
\centering
\lstCode
\begin{lstlisting}[
    label={lst:vector_reserve_pattern_b},
    caption={Optimization example for Pattern B.},
    language=code
]
result_entities.reserve(result_entities.size() + source_entities.size());
absl::c_move(source_entities, std::back_inserter(result_entities));
\end{lstlisting}
\end{figure}

\noindent Clang-Tidy fails here because the insertion is encapsulated inside an algorithm wrapper and iterator adapter, leaving no explicit loop in the local AST. \name{} is able to recognize the element transfer semantics and recommends reserving.

\niparagraph{Pattern C: Iterator-based Assignment on Custom Containers}
In production, developers frequently populate proprietary containers, such as a protocol buffers in a \texttt{RepeatedPtrField}. In the example below, \texttt{cache\_value} is a protocol buffer.

\begin{figure}[H]
\centering
\lstCode
\begin{lstlisting}[
    label={lst:vector_reserve_pattern_c},
    caption={Optimization example for Pattern C.},
    language=code
]
cache_value.mutable_entities()->Reserve(result.entities.size());
*cache_value.mutable_entities() = {result.entities.begin(), result.entities.end()};
\end{lstlisting}
\end{figure}

\noindent Clang-Tidy fails because its analysis rules are hardcoded for standard library types (mainly \texttt{std::vector}). It cannot reason about custom APIs or allocations triggered implicitly via assignment operators and initializer lists. By using an LLM, \name{} can leverage learned API signatures to suggest the custom container's \texttt{Reserve()} method.

\niparagraph{Pattern D: Generic Template Sizing \& Iterators}
Generic, templated code is commonly used to write reusable copy utilities.

\begin{figure}[H]
\centering
\lstCode
\begin{lstlisting}[
    label={lst:vector_reserve_pattern_d},
    caption={Optimization example for Pattern D.},
    language=code
]
template <typename Range>
std::vector<Entity> Copy(const Range& source) {
  std::vector<Entity> destination;
  destination.reserve(source.size());
  destination.insert(destination.end(), source.begin(), source.end());
  return destination;
}
\end{lstlisting}
\end{figure}

\noindent At analysis time, Clang-Tidy cannot resolve the exact type of the template parameter \texttt{Range}. It systematically ignores uninstantiated parameters to avoid risking broken code or generating complex fallback rule paths. LLMs are able to infer the range properties of abstract types and apply block insertions.

\niparagraph{Pattern E: Inter-Procedural / Multi-file Size Mappings}
In complex systems, sizing information is often decoupled and evaluated across functional or compilation unit boundaries.

\begin{figure}[H]
\centering
\lstCode
\begin{lstlisting}[
    label={lst:vector_reserve_pattern_e},
    caption={Optimization example for Pattern E.},
    language=code
]
// Sub-optimal Implementation
void PopulateStats(const std::vector<Entity>& src, std::vector<Entity>& dest) {
  for (const auto& item : src) {
    if (item.isValid) { dest.push_back(item); }
  }
}

// Optimized Implementation
void PopulateStats(const std::vector<Entity>& src, const CollectionStats& stats, std::vector<Entity>& dest) {
  dest.reserve(stats.total_size);
  for (const auto& item : src) {
    if (item.isValid) { dest.push_back(item); }
  }
}
\end{lstlisting}
\end{figure}

\noindent Clang-Tidy is limited to single-file compilation units and cannot trace size constraints spanning inter-procedural boundaries. Additionally, conditional loops prevent direct mapping of \texttt{src.size()}. With an LLM, \name{} can map dependencies across compilation boundaries to recommend a \texttt{.reserve()} utilizing the pre-calculated size member (\texttt{stats.total\_size}).

\niparagraph{Pattern F: Custom Pointer Array Buffers}
This pattern involves custom logging or cache buffers that store object pointers instead of copying full structures. In this case, \texttt{buffer} is a custom, dynamically sized data structure that holds a set of pointers, which can be added to it using \texttt{Add}.

\begin{figure}[H]
\centering
\lstCode
\begin{lstlisting}[
    label={lst:vector_reserve_pattern_f},
    caption={Optimization example for Pattern F.},
    language=code
]
buffer.Reserve(source.size());
for (auto& item : source) { buffer.Add(&item); }
\end{lstlisting}
\end{figure}

\noindent Similar to Pattern B, Clang-Tidy ignores this loop because the container is a custom class and uses a non-standard insertion method (\texttt{.Add()}). Clang-Tidy has no semantic awareness that \texttt{.Add()} triggers underlying array expansions. LLMs can identify the custom buffer's memory semantics and suggest a \texttt{Reserve()} call.

\end{document}